\documentstyle[11pt,psfig]{article}
\voffset = - 1.7 cm
\hoffset = - 1.3 cm
\newcommand{\blankline}{\vskip .3cm}
\newcommand{\f}{\begin{equation}}
\newcommand{\ff}{\end{equation}}

\newcommand{\bra}[1]{\langle #1 \mid}
\newcommand{\ket}[1]{\mid #1 \rangle}
\newcommand{\braket}[2]{\langle #1 \mid #2 \rangle}
\newcommand{\pic}[5]{\raisebox{#3pt}{
\hspace{#4pt}\psfig{file=#1.ps,height=#2pt,silent=}\hspace{#5pt}}}
\newcommand{\kd}[1]{\mathchoice{
\pic{#1}{20}{-6}{-1}{2}}{
\pic{#1}{11}{-1}{-3}{2}}{
\pic{#1}{9}{-2}{-3}{1}}{
\pic{#1}{7}{-1}{-1}{0}}}

\begin{document}
\rightline{\Large CGPG-95/12-3}
\centerline{\LARGE  Quantum deformation of quantum gravity}
\blankline
\rm
\vskip.3cm
\centerline{Seth Major${}^*$ and Lee Smolin${}^\dagger$ }
\blankline
\centerline{\it  Center for Gravitational Physics and Geometry}
\centerline{\it Department of Physics, The Pennsylvania State
University}
\centerline{\it University Park, PA, USA 16802}
\blankline
\centerline{8 December 1995 }
\blankline
\centerline{ABSTRACT}
We describe a deformation of the observable algebra of quantum gravity
in which the loop algebra is extended to framed loops.   This allows an
alternative nonperturbative quantization which is suitable for  describing
a phase of quantum gravity characterized by states which are normalizable
in the measure of Chern-Simons theory. The  deformation parameter, $q$, is
$e^{ i \hbar^2 G^2 \Lambda /6}$, where $\Lambda$ is the cosmological
constant.  The  Mandelstam identities are extended to a set of relations
which are governed by the Kauffman bracket so that the spin network basis
is deformed to a basis of $SU(2)_q$ spin networks.  Corrections to the actions
of operators in non-perturbative quantum gravity may be readily computed
using recoupling theory; the example of the area observable is treated here.
Finally, eigenstates  of the q-deformed Wilson loops are constructed, which
may make possible the construction of a q-deformed connection representation
through an inverse transform.
\noindent
\blankline
internet addresses:  ${}^*$seth@phys.psu.edu
${}^\dagger$smolin@phys.psu.edu

\section{Introduction}

In the past five years  a number of striking
consequences of diffeomorphism invariance have
emerged in the non-perturbative approach to quantum gravity
based on the loop representation
(\cite{carloleeloop} - \cite{troy}
For reviews
see \cite{reviews,ls-review}).
One of these provides a {\it basis}
of spatially diffeomorphism invariant states
labeled by diffeomorphism equivalence
classes of embedding of
spin networks\cite{spinnetcl,volumecl}.\footnote{Note that
the elements of the basis
are differentiated by labels attached to vertices of valence higher
than three.  This may be done according to Figure 5.}
In this context, a spin network is a graph with edges labeled
by representations of $SU(2)$ and
vertices labeled by the ways that the edge representations
may be combined into a singlet.  This concept of a spin
network was first introduced by
Penrose\cite{roger-sn} in work
on the four color problem.  Later, he applied spin networks
to a combinatorial construction of
geometry.\footnote{Penrose restricted attention to trivalent spin
networks, which are especially simple in that the
vertices are unique and hence require no separate labels.}  The
concept independently reappeared in
lattice gauge theory where spin networks
label bases\cite{ks,fk} of states, a property equally
useful in non-perturbative quantum gravity where spin
networks were introduced following the discovery
that this basis diagonalizes
two interesting classes of observable, areas and
volumes\cite{volumecl}.

This work constructs one natural extension of
these results to a class of theories in which the
role of the spin networks is replaced by a closely related set of
combinatorial and topological networks called
{\it quantum spin networks} or q-spin nets.
These structures have emerged in
the investigation of topological quantum
field theory, and play
a key role in elucidating the connection between
Chern-Simons theory and the
Kauffman bracket\cite{lou-qnet}.  Closely
related to the topological and algebraic
structures which underlie
conformal field theory\cite{atiyah,segal,moreseiberg} and hence perturbative
string theory, quantum spin networks turn out also to be central
to category-theoretic foundations
of topological quantum field theory in three and
four dimensions
\cite{ctqft}.

The need for a deformation of the loop algebra is evident in
Chern-Simons theory as expectation values of
loop observables
\f
{\cal K}[\gamma ] = \langle \hat{T}[\gamma ] \rangle_{CS}=
\int d\mu [A] \exp \left( { k \over 4 \pi } S_{CS} \right) T_\gamma[A],
\nonumber
\ff
where $T_\gamma[A]$
is the Wilson loop of the connection $A$ around $\gamma$,
are not defined.  There exist divergences which can be
removed only if the loops are
framed\cite{witten-cs}.  Once this is done,
the integral defines the Kauffman bracket, which is a diffeomorphism
invariant function of the embeddings of framed loops\cite{lou-qnet}.
The expectation values of loops
define a set of identities which extends the
Mandelstam identities satisfied by Wilson loop observables.
This means that the measure $d\mu [A]$, cannot be one of the
diffeomorphism invariant
measures\cite{almeasure,baezmeasure,MU} constructed in studies
of quantum gravity in terms of elements of the completion
$\overline{{\cal A }/ {\cal G}}$.

This is relevant for quantum gravity because of the
Kodama state\cite{kodamastate}
\f
\Psi_{CS}[A ] = \exp{ \left[ {3 \over \lambda} S_{CS}(A) \right]}
\nonumber
\ff
where $\lambda= G^2 \Lambda$ is the dimensionless cosmological
constant and $S_{CS}(A)$ is the Chern-Simons invariant of the
left handed Ashtekar-Sen connection $A$.
This state is one of the few
explicit solutions to the constraints of quantum
gravity in the connection representation.  Furthermore,
for small $\lambda$, it may be interpreted as a
semi-classical states associated with
De-Sitter spacetime\cite{cstime}.   It is then interesting to
hypothesize that
this state gives a non-perturbative description of the
vacuum state in the presence of the
cosmological constant.  To investigate this hypothesis we may study
excitations of the Kodama state, of the form
\f
\Psi [A, \phi ] =\Psi_{CS} [A ] \Xi [A , \phi ]
\label{swm}
\ff
where $\phi$ is a matter degree of freedom.  Among these are the states
\f
\Psi [A] = \Psi_{CS} [A ] T_\gamma[A].
\label{kodamastate}
\ff

Alternatively,
in the presence of boundaries, the Chern-Simons state seems to
define a  sector of the theory in the loop
representation, of states\cite{lslinking,lsbekenstein},
\f
\Psi_\rho [\gamma ] = \int d\mu [A,a] \Psi_{CS} [A ] T_\gamma[A] \rho [a]
\nonumber
\ff
where $a$ is a suitably defined
holomorphic part of the pull back of the Ashtekar
connection to the boundary and $\rho [a]$ is a state of the
Chern-Simons theory of the boundary.  These states may be
sufficient to span the physical state space as they saturate
the Bekenstein bound when the boundary has a fixed,
finite area \cite{lslinking,lsbekenstein}.

For these reasons, it seems likely that in the presence of either a
cosmological constant or appropriate boundary conditions
quantum gravity will be formulated in terms of the Kodama state.
In the loop representation, however, expressions
such as Eq. (1) are not defined unless the loops are framed. Thus,
we construct an extension of the loop representation to include
states which are functionals of framed loops.
One way to do this is to construct an extension of the loop
algebra.   This is the main goal of this paper.
We shall see that there is a natural modification of
the loop algebra involving framed loops and an extended set of identities
that combine the
Mandelstam identities with the relations satisfied by the
Kauffman bracket.  The
resulting algebra has a representation which is spanned by a basis
labeled by $q$-deformed spin networks.

The deformation parameter, $q$
\begin{equation}
q=e^{ i \pi / r}
\end{equation}
with $r=k+2$ arises through the dependence of the Kodama state on
the cosmological constant.  With Newton's constant $G$ and $\hbar$
the coupling constant of Eq. (1) is
\f
k = {6 \pi \over \hbar^2 G^2 \Lambda}  +\alpha. \label{defk}
\ff
where $\Lambda$ and $\alpha$ are, respectively, the cosmological constant
and the value of a $CP$ breaking
phase coming from a $\int F\wedge F$ term in the action.
The  definition of Eq. (\ref{defk}) implies that the cosmological constant
must take on discrete values \cite{lslinking}.
In addition, the limit in which
$k \rightarrow \infty$ removes the effects of framing
so that quantum spin networks return to ordinary spin
networks.   As a result, the algebra we describe here
may be thought of as a deformation in $\hbar^2 \Lambda$
of the classical
loop algebra, which incorporates framing of loops as a quantum
effect which goes away in the limit $\hbar \rightarrow 0$.

This may seem a bit peculiar, as the cosmological
constant is usually expected to only influence the large scale,
and to only affect the theory at the level of dynamics.  However, since
the representation of the
ordinary loop algebra leads to the spin network basis, the deformation in
$\hbar^2 \Lambda$ must be taken into account in the kinematical
algebra of the theory.  Indeed, it is
common in quantum field theory for the kinematical state space
of the theory to be modified to incorporate dynamics.
For instance, one discovers in rigorous
studies of  $\phi^4$-theory in $2$ and $3$
dimensions that, by Haag's theorem, we cannot implement dynamics with the Fock
space quantization of the associated free field theory.
In addition, the structures of the kinematical state
spaces know about the mass $m$, which is the parameter of
highest dimension in scalar field theory, as $\Lambda$ is the
parameter of highest dimension in gravitational theory.
We then conjecture that the cosmological constant may play
an analogous role in quantum gravity, and so requires a deformation
of the observable algebra and representation of the quantum
theory at the kinematical level.

Our goal is to investigate this
conjecture by showing that there is a suitable
deformation of the algebra which
yields cosmological constant corrections to physical observables.
Thus, our basic hypothesis
is that a sector, or phase, of quantum gravity, given
by excitations of the Kodama state, is
the physical phase in the presence of a cosmological
constant. We call this the  ``Kodama phase'' of
quantum gravity.  In this phase, purely
quantum effects add a degree of
freedom to the loops which counts
the twisting of loops.  It is mathematically described with
framed loops and quantum spin networks.

In the next section
we define a formal algebra of framed loops.
In Section 3 we describe a representation of this
algebra in terms of suitable functionals of framed loops
and show that it
has a basis given by the embeddings of the $q$-deformed spin
networks.  Sections 4 and 5 describe, respectively, the extension
of the algebra to deformations of $T^1$ and $T^2$ operators.  The
latter allows us to define and compute eigenvalues of the
$q$-deformed area operator.  Eigenstates of the
deformed Wilson loops are constructed in Section 6, and the
paper ends with comments on directions for future work.

In closing, we warn the reader that the considerations of this paper
are mathematically heuristic.  However,
the mathematical structures we use here are not new; indeed
this paper may be read as a proposal to apply the
mathematical structures of
Kauffman\cite{lou-qnet} and
Kauffman and Lins \cite{KL} to quantum gravity.  We
establish physical arguments for the application to quantum
gravity of these mathematical structures.
Interesting questions such as whether there exist measures
on $\overline{{\cal A}/{\cal G}}$ \cite{almeasure,baezmeasure}
associated to framed loops
or a useful $q$-deformation of the notion of a
connection are not treated here.
Finally, we mention that work is underway in
collaboration with R. Borissov to compute the
action of the deformations of operators such as the volume and
$H =  \int \sqrt{-C}$, where $C$ is the Hamiltonian constraint of
quantum gravity\cite{inpreparation}.

\section{The framed commutative loop algebra}

We preface this paper with two remarks.  First, our
hypothesis has an important
consequence for diffeomorphism
invariant regularization procedures.
As is described in \cite{ls-review}, naive operator products
derive meaning through a limit procedure in which loops -
introduced to make point split operators gauge invariant - are
shrunk to points.  These limits are outside the topology defined by the
diffeomorphism invariant states; values of diffeomorphism invariant
states on the ``shrunk loops" differ { \it discontinuously} from
values on finite loops.  These limits require
new topologies which are external to the
structure of diffeomorphism invariant state spaces.

The standard definition of these limits assumes they are state independent.
However it is clear that this is not always true.  This is shown,
for example, by
the behavior of the loop operators in the limit that a loop is
shrunk to a point.  According to the standard
definitions of the loop representation,
if $\beta^\delta $
is a one parameter family of loops such that, in some background
euclidean metric, each is a circle of radius $\delta$ then,
under standard definitions, the limit
\f
\lim_{\delta \rightarrow 0} \, \bra{\alpha }\hat{T} [\beta^\delta  ]
= - 2 \bra{\alpha}
\ff
is independent of the relationship between the loops $\beta^\delta$
and the loop $\alpha$ (Here, we use a choice of trace on the group
corresponding to
``binor notation;'' see Section 2.2).  On the other hand, in the
presence of the Kodama state the actions of loop operators are
given by the path integral of Chern-Simons theory \cite{witten-cs}
or, equivalently, by the Kauffman bracket.  In the limit that loops
are shrunk down the effect of the loop operator differs from Eq. (8).
Instead we have
\f
\lim_{\delta \rightarrow 0} \int d\mu [A] \Psi_{CS} [A ]
T_{\beta^\delta} [A]
T_\alpha [A] = \left( -q - q^{-1} \right) \int d\mu [A] \Psi_{CS} [A ]
T_\alpha [A]
\ff
for loops $\beta^\delta$ which have vanishing
linking number with $\alpha$.  If the loops are linked then the
limit depends on linking number as well.  Thus, in
defining a new loop representation to describe the Kodama state
and its excitations, the standard assumptions
made in the construction of the diffeomorphism invariant regularization
procedures must be extended.
Happily, these examples suggest how to modify the usual procedure.
It is natural to require that, instead of the naive limits such as Eq. (8),
{\it loops, in the limit of a regularization procedure,
are governed by Kauffman bracket relations.}
This requirement holds, by definition, for all states
of the form of Eq. (4).    We call one parameter
families of loops which have point limits  ``sloops" for ``shrinking loops.''
This hypothesis, which determines the combinatorics of sloops,  will be
denoted the ``sloop hypothesis.''

The second remark is that we could define the deformed
loop algebra directly in terms of its action on the q-spin network basis.
To do this one only needs to compute the action of  loop
operators in the spin network basis and then deform
that action to a basis labeled by
$q$-deformed spin networks.  While the end result is equivalent to
what we do here, we take the less direct course as it is
convenient to have deformed
equivalence classes of loops in order to verify relations and
perform calculations.

\subsection{The basic strategy}

Our goal is to construct a framed loop
algebra, ${ \cal LA}^f$.  We first define a free
complex vector space, ${ \cal FL}^f$ of formal
linear combinations of framed multiloops.  On
this space we define a product and an
equivalence class generated by a list
of relations which extend and generalize
the Mandelstam relations of standard loop observables.
These relations realize the hypothesis
that the Kauffman bracket relations hold for sloops.

The product on ${\cal LA}^f$, $\alpha^f \cup \beta^f $,
will be commutative and associative.  This extends the usual
commutative algebra of $SU(2)$ Wilson loop observables allowing
us to define a deformed algebra of framed loop operators
denoted $\hat{T}_q \left[\alpha\right]$ such
that
\f
\hat{T}_q [\alpha] \hat{T}_q [\beta] =\hat{T}_q[\alpha \cup \beta]
\nonumber
\ff
Once the algebra is defined we find the
representation which is a deformation of the usual
loop representation.  Finally, higher order $\hat{T}$ operators
are constructed in this representation.

\subsection{Technical note: binor diagramatics}

We define ${ \cal FL}^f$ in terms of its so-called
``binor representation.''   An element of the
vector space is indicated by a two dimensional diagram, which
is called the framed loop diagram of $\alpha^f$,
indicated $P (\alpha^f)$.  The loop in the
spatial manifold is indicated by labeling the edges of the diagram.
This diagrammatic notation is defined so that
the limit in which the deformation parameter $q \rightarrow 1$ takes us
to algebra of $SU(2)$ Wilson loop observables, expressed in
a diagrammatic notation due to Penrose called the binor notation.
The binor notation has built into it two sign rules which come
into the correspondence between the diagram of a loop
$P(\alpha )$ and the Wilson loop functionals.
These correspond to a definition of the trace of a parallel
transport so that $Tr[1]=-2$ together with
an assignment of $-1$ to every crossing.
This notation has the important advantage that it is local
and topologically invariant in the two dimensional plane in which
the diagrams live.  This greatly simplifies calculations.
The Mandelstam identities become
\f
\kd{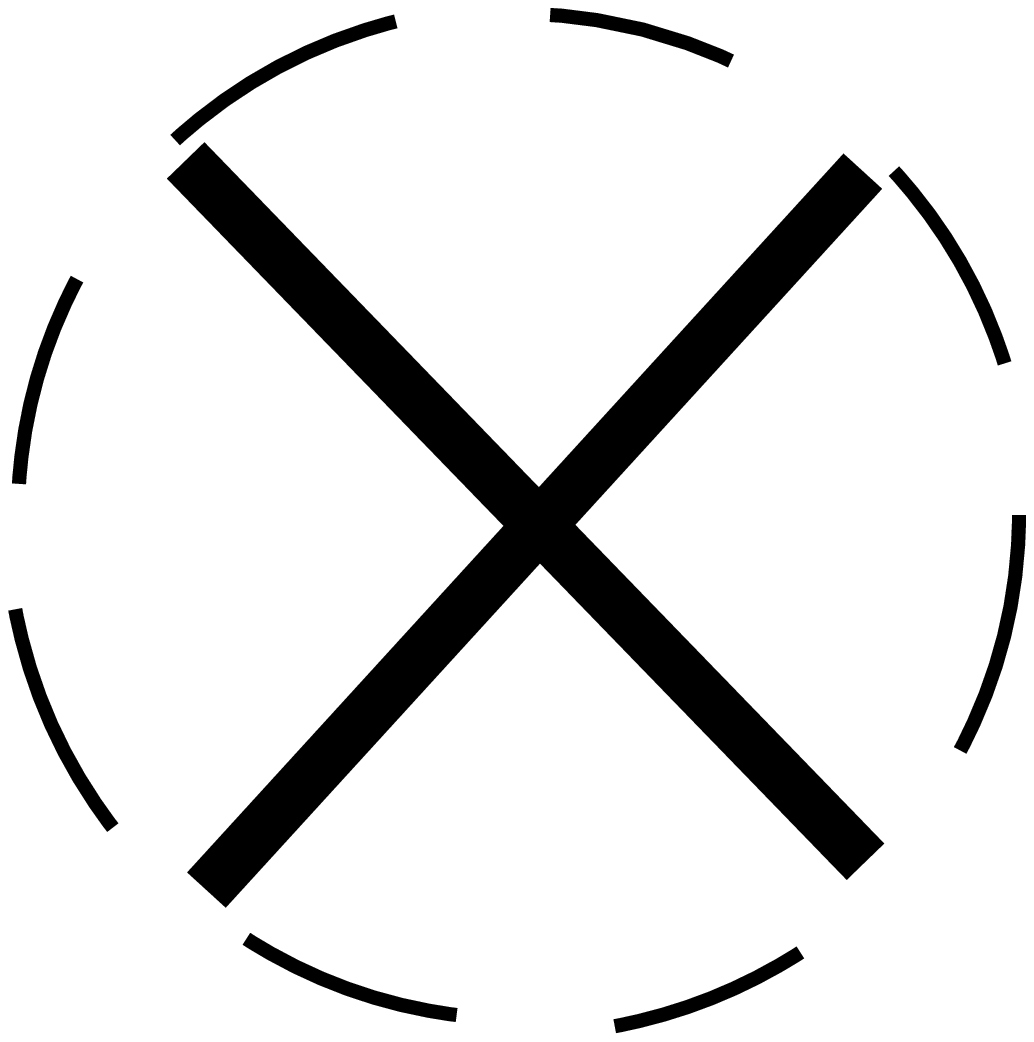} + \kd{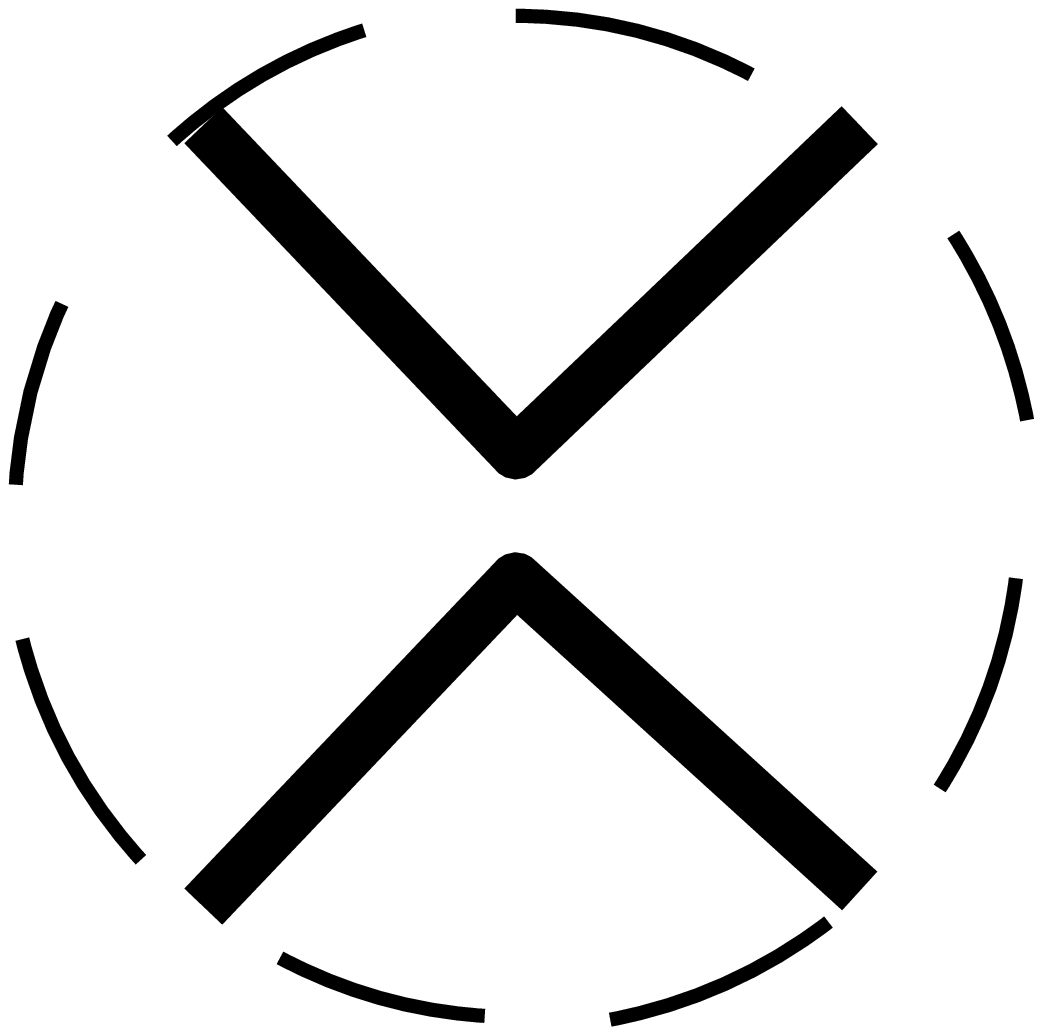} + \kd{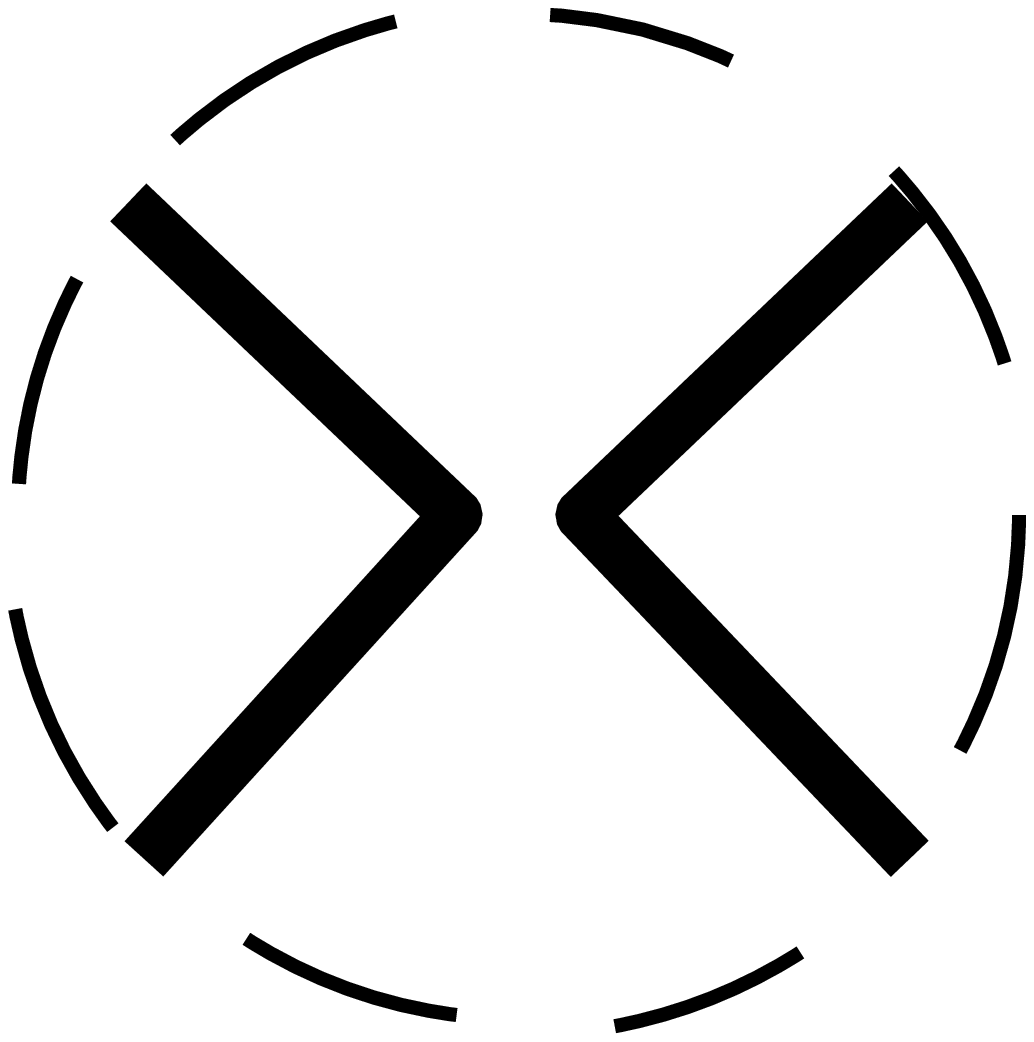} = 0.
\label{mand}
\ff
(Diagrams such as these circled by a dashed line represent changes occurring
at a point; the parallel transport along edges inside dashed circles
are trivial.) Symmetrizations over spinor indices
of elements in the
connection representation are (due to the added sign) represented by
anti-symmetrizations over multiloops in the binor diagram
$P(\alpha )$.

We can express the deformation in
terms of a deformation parameter $A$ such that
\f
A^2 =q .
\ff
The usual binor representation
is then recovered by taking the limit in which
$A \rightarrow -1 $.

\subsection{Framed loops }

The motivation for defining framed loops arises from
defining operator products through regularization procedures
for the Kodama phase.
In these regularization procedures
new loops are introduced
to connect points that are ``split apart'' in
operator products. The resulting operators are defined as
limits in which these loops are shrunk down.
Ambiguities in these limits, due to the Chern-Simons factor, may be
resolved with a finite amount of topological information.
This can be encoded in framing.
As shown in \cite{witten-cs}, expectation values such as in Eq. (1)
depend on an integer -- the self-linking number of the loop.
However, to fully define a framed loop it will not be sufficient to
append a self-linking number.  Additional ambiguities arise when the
loops intersect. To resolve these, the definition
of a framed loop will involve additional information associated with each
intersection point.
Let us begin with the definition of non-intersecting framed paths.

An individual framed path, denoted by $\pi^f$
is a path $\pi:I \rightarrow \Sigma$, with a direction field
associated to every point of the path - the ``framing.'' Framing
can be seen as a direction in a plane perpendicular to the tangent
vector $\dot{\pi}^a(s)$ of $\pi(s)$ i.e.
$\pi^f:I \rightarrow \Sigma \times S^1$. Framed loops,
denoted $\alpha^f, \beta^f , \gamma^f , \ldots$, are closed paths,
$\alpha(0)=\alpha(1)$ with a continuous direction field,
denoting the frame of $\alpha^f$ as $\theta_\alpha$, $\theta_\alpha
(0) = \theta_\alpha(1)$.
A framed multiloop, which following the original loop formulation
\cite{carloleeloop}, will be also
denoted by greek letters, is a set of individual framed loops.
The identity for framed loops is the constant map, $e$, with $L(e)=0$.

The framing is defined modulo smooth deformations
of the direction field.  As such
all that is relevant to define the framing of
a non-self-intersecting loop is the self-linking number, which is  the
number of times the direction field wraps around the loop.  More explicitly,
the  self-linking number $L(\gamma^f)$
of a framed loop, $\gamma^f$,
is defined in terms of a linking number.

The linking number
$L(\gamma , \beta )$ of two distinct non-intersecting loops
$\gamma$ and $\beta$ may be expressed in terms of a two dimensional
projection, or diagram, of the loops.  These loops, given an orientation, have
linking number
\begin{equation}
L(\gamma, \beta) = {1 \over 2 } \sum_c \epsilon(c),
\label{link}
\end{equation}
where $c$ sums over all the crossings in the diagram and
$\epsilon=1$ is for over crossings,$\kd{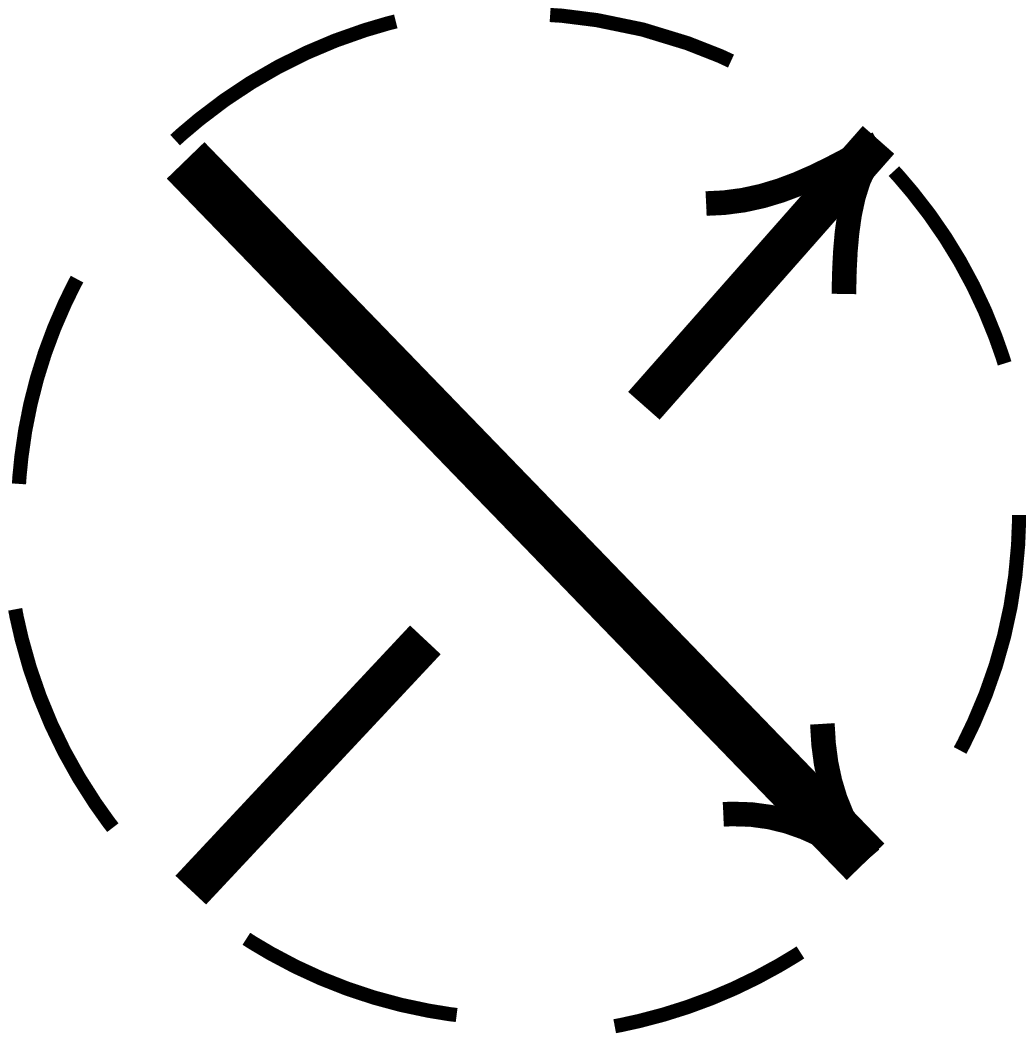}$, and $\epsilon=-1$ for
under crossings, $\kd{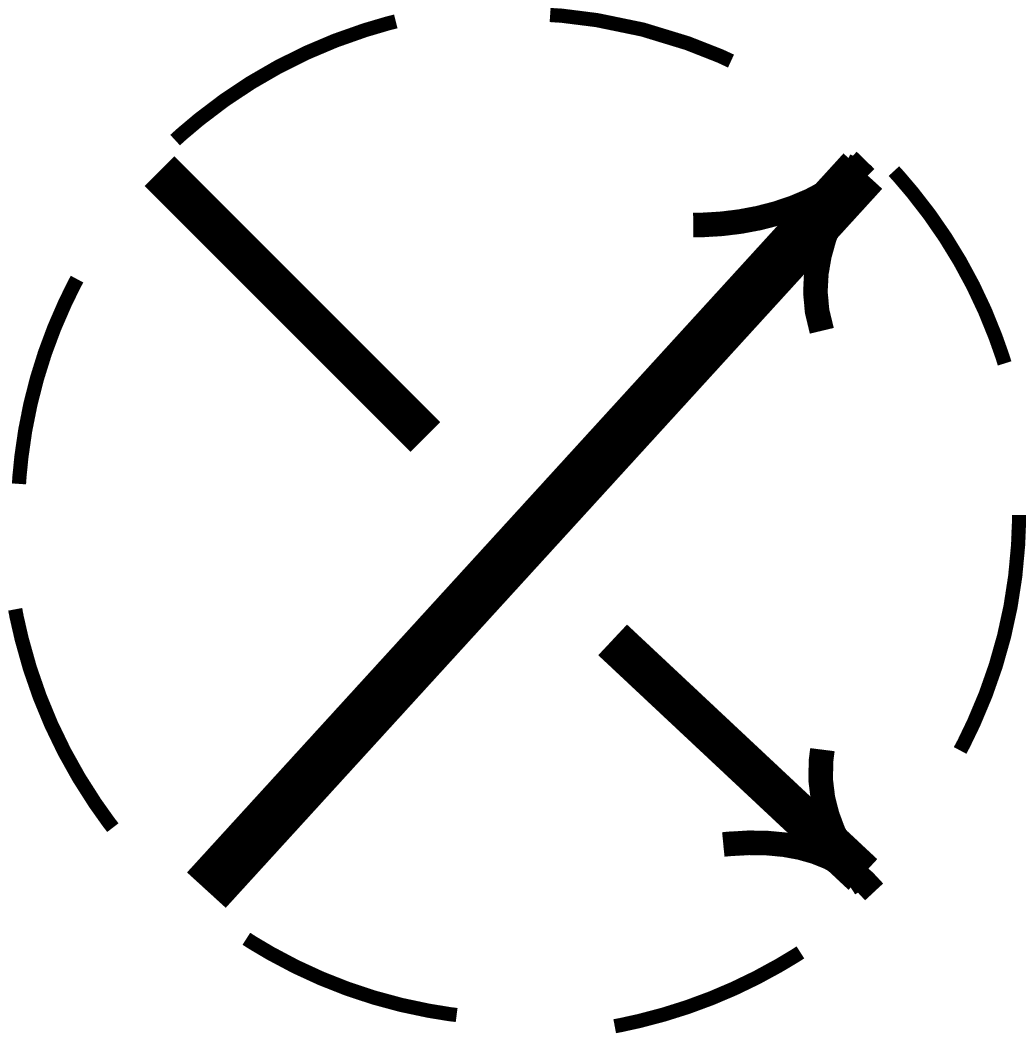}$. This is
a diffeomorphism invariant quantity.
\begin{figure}
\begin{tabular*}{\textwidth}{c@{\extracolsep{\fill}}c@{\extracolsep{\fill}}
c@{\extracolsep{\fill}}}
\pic{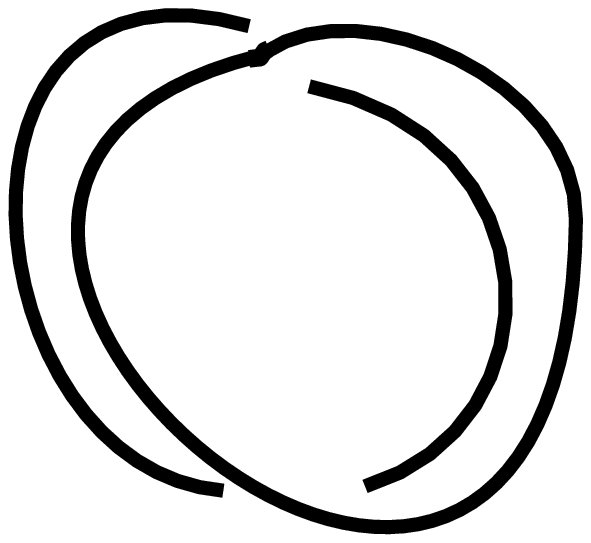}{50}{5}{0}{0}&
\pic{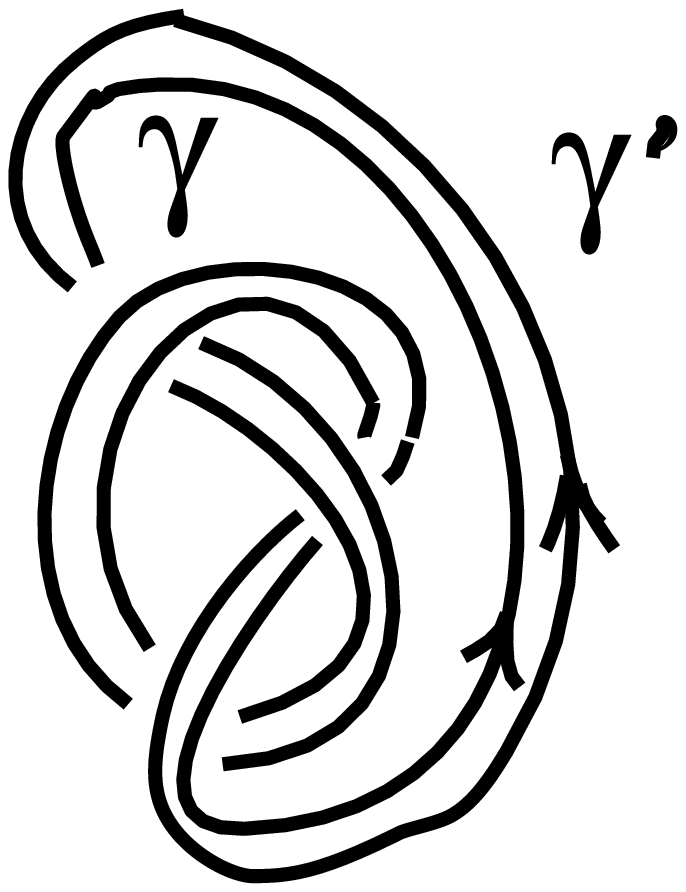}{60}{5}{0}{0}&
\pic{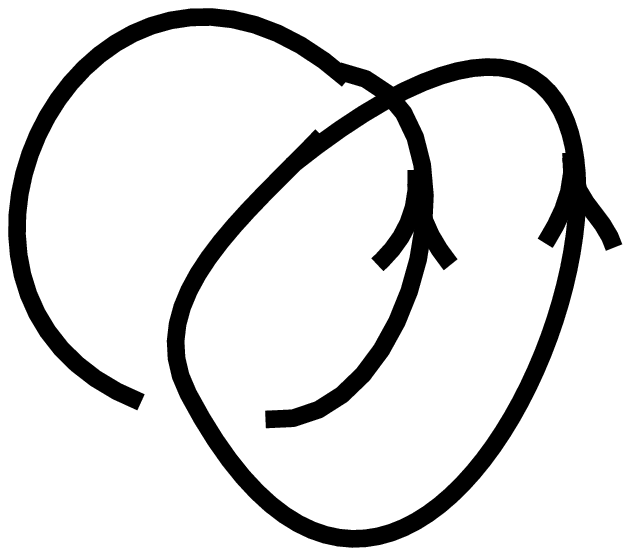}{50}{5}{0}{0} \\
a & b & c \\
\end{tabular*}
\caption{Examples of framing: (a.) Two unlinked unknots, $L=0$  (b.)
A knot with a direction field in the plane of the diagram - ``blackboard
framing'' - giving a linking number $-2$ between
the knot $\gamma$ and its frame $\gamma'$ (c.) A
pair of intersecting unknots with linking number $L=B$ [See Eq.
(\ref{blink})].}
\label{linkfig}
\end{figure}

The self-linking
number can be computed using the framing direction field.
In some background metric, a framed loop $\gamma$ is
displaced an
infinitesimal distance in the direction field to obtain
another loop $\gamma'$.  Once an orientation is given,
the self-linking number $L(\gamma)$ is the linking number between these
two loops $L(\gamma, \gamma')$, given by Eq. (\ref{link}).
An example is given in Fig. (\ref{linkfig}b).

The inverse of a
framed loop is defined as
reversing the tangent vector of the loop, keeping the self-linking number
fixed so that the direction field ``reverses'' or, is mapped to
the antipodal point of $S^1$.

When there are intersections, additional
information is needed to define
a ``framed loop.'' This can be seen if we regard all intersections and
overlapping paths as {\it limits points} of sequences of non-intersecting
loops.  These sequences can approach the intersection in
a variety of ways.  Framing encodes
topological information in the limits.  For example,
two loop segments can, in the limit, touch ``from the top", or ``from the
bottom.''  These two cases may be diagrammatically represented by
$\kd{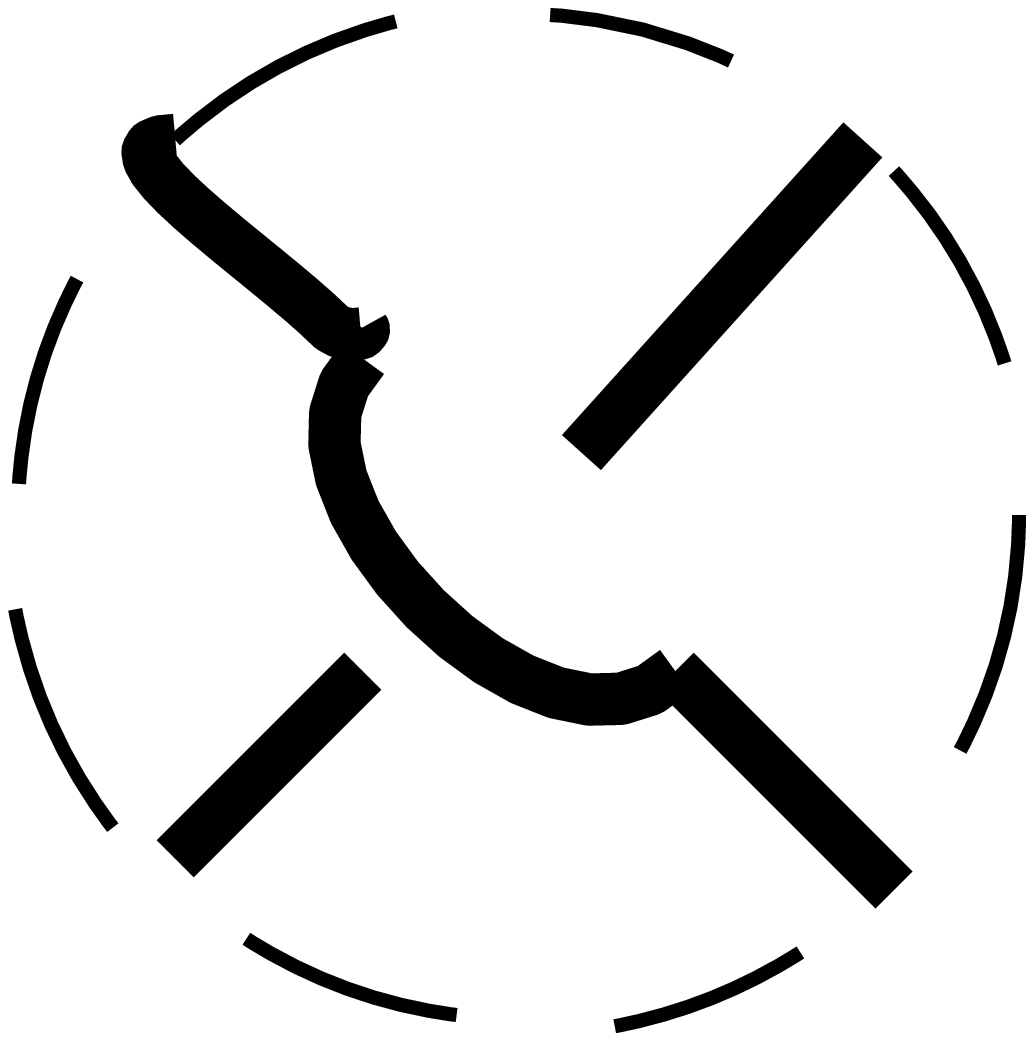}$ and $\kd{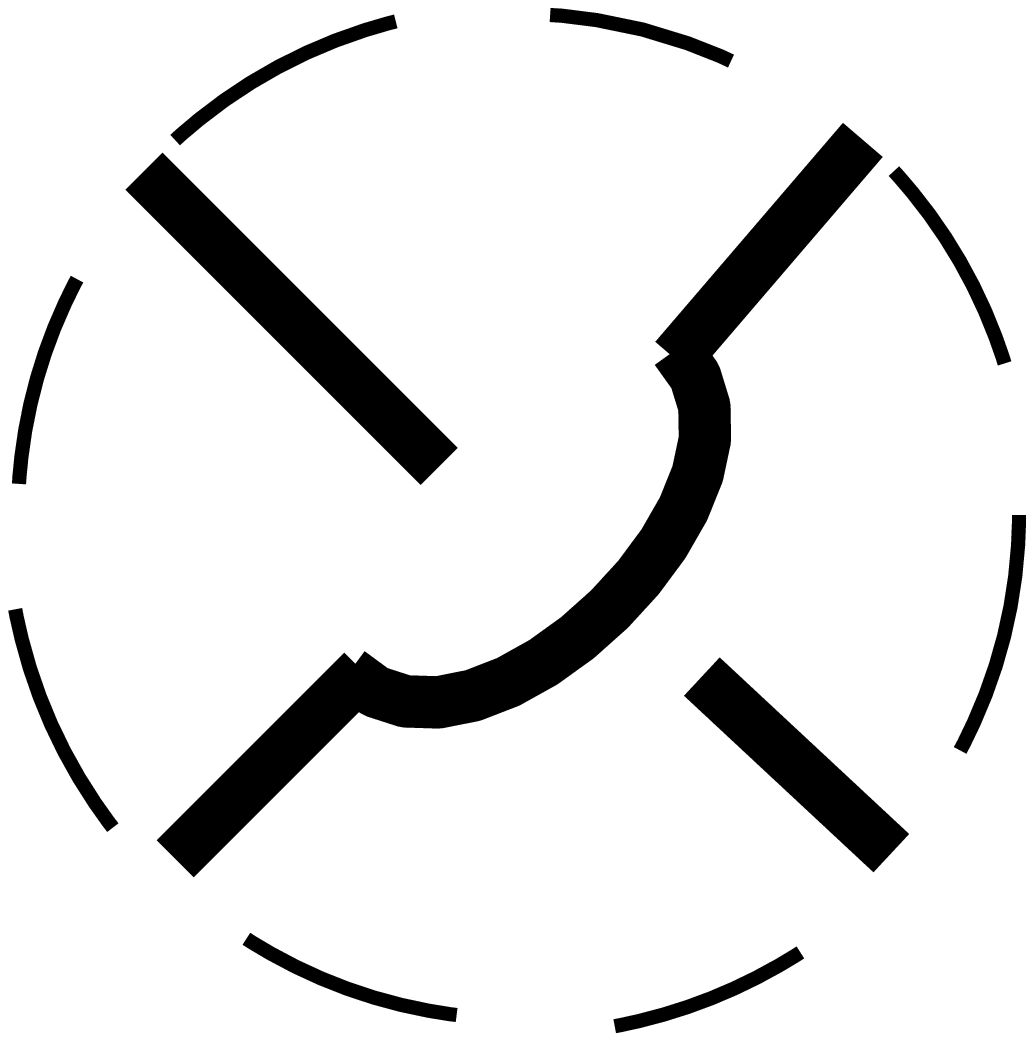}$ which we call
``touching from the top'' and ``touching from the bottom,''
respectively,  and may be thought of as two, distinct results of
a limit of a regularization procedure in which
loops are brought together.
More generally, we can think of the space of
loops with intersections as the completion of the space of non-intersecting
loops. These intersections $\kd{uptouch}$ and
$\kd{dntouch}$ represent distinct points in this space.\footnote{We thank
Carlo Rovelli for suggesting this perspective.}

As the self-linking numbers of the loops are defined only up
to arbitrary smooth deformations of the direction field,
they play no role
at an intersection.   Instead, in a neighborhood of the intersection the
direction fields of each loop may be deformed smoothly so that
the fields lie in the plane
formed by the tangent vectors at the intersection.  Given this
freedom we can define the intersections in ``blackboard'' framing
in which the plane of projection is determined by the tangent
vectors. All
intersections may then be expressed as a framing factor
times one of these intersections.  To summarize, in terms of limits
of non-intersecting loops there are precisely two,
distinct ways that two framed loops can meet at a point, given by
$\kd{uptouch}$ and $\kd{dntouch}$.  For the purposes of
the quantum theory we take these to span a two dimensional
space of possible states associated with the intersection.

Given $\kd{uptouch}$ and $\kd{dntouch}$ as basis
elements for the different states associated with the
intersection, we may define linear combinations of them that
correspond to intermediate cases.
These will be of the form
\f
\kd{cross}_z = z \kd{uptouch} + z' \kd{dntouch}
\ff
where $z$ and $z'$ are complex numbers.  Of particular interest
is a combination defined by $z=z'=B$
\f
\kd{cross} = B \left (\kd{uptouch} +  \kd{dntouch} \right )
\label{bmandle}
\ff
in which the coefficient $B$ will be chosen below so that these
kinds of intersections
satisfy the ordinary Mandelstam identities.

Loops can not only pass through each other at points
of intersections, there can be ``exchanges of parallel transport"
such as in $\kd{collision}$
and $\kd{wedges}$.  In the case of
ordinary loops these are related to the
unique intersection state $\kd{cross}$ by the Mandelstam
identity Eq. (11).  For framed loops we define
$\kd{collision}$ to be the limit of a sequence in which the two loops meet
at a point. The other case $\kd{wedges}$ is defined similarly.
Other kinds of intersections are defined in terms of
these by the equivalence relation that will
be defined in the next subsection.  Furthermore,
these equivalence relations will leave
us, as in the case of ordinary loops, with only two independent states
associated with the routings and framings of a simple intersection,
$\kd{cross}$.
These will be defined so that
Kauffman bracket relations are recovered for sloops.

It is useful to extend the notion of linking numbers to cases involving
intersections.  Since each such case is defined as a limit
of a sequence of non-intersecting loops it is straightforward in these
cases to define the linking numbers in terms of these sequences. Generally
the linking numbers of intersecting loops are found deforming the loops
slightly in the direction inverse to the limit that defined the intersection.
In the case of an ``touch from the top" we  deform as
\begin{equation}
\kd{uptouch} \rightarrow \kd{ovrcross},
\end{equation}
and then compute the linking number.   The linking number associated with
other linear combinations are then defined by the condition
\f
L(\alpha^f , \beta^f + \gamma^f) = L(\alpha^f , \beta^f ) +
L(\alpha^f , \gamma^f)
\nonumber
\ff
For example,
\begin{eqnarray}
L \left( \kd{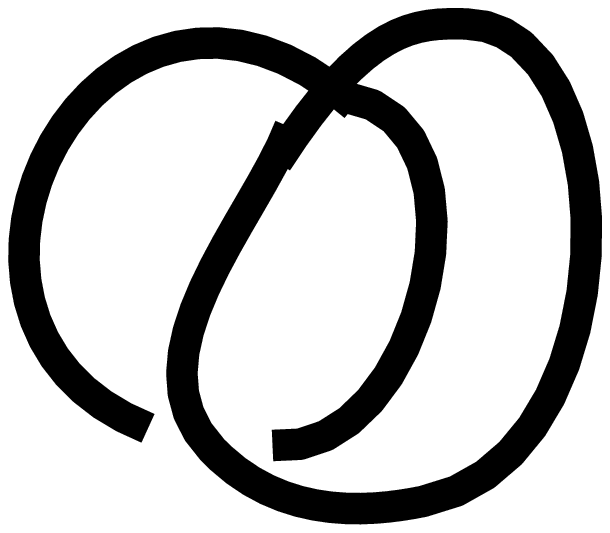} \right) &=& B \left[ L \left( \kd{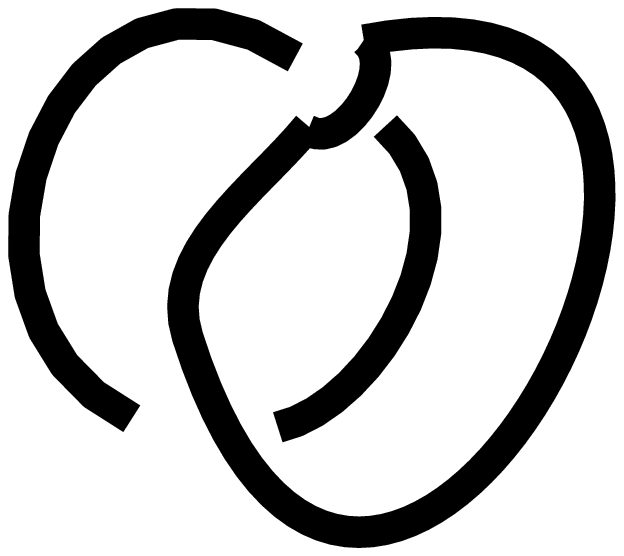}
\right)+ L \left(\kd{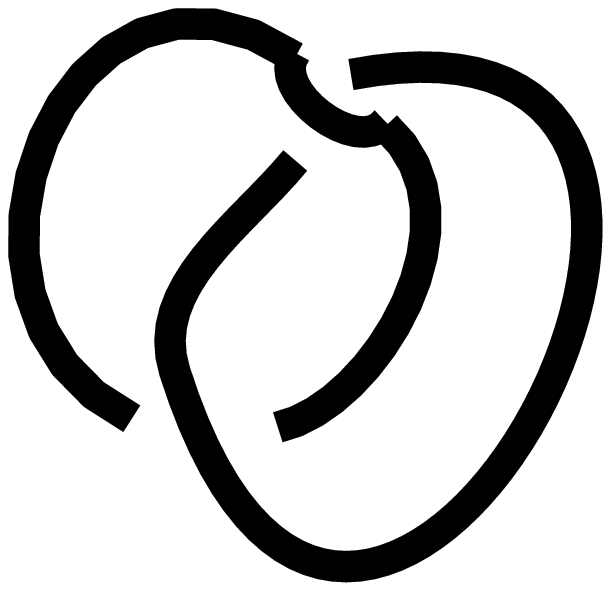} \right)\right]
\nonumber \\
&=& B \label{blink}
\end{eqnarray}
Arbitrary intersection points, at which any number of
paths meet, may be described with this principle of completing
the space of non-intersecting loops. This is done separately,
in \cite{BMS}.

Finally, it is useful to extend the usual definition of the product
(or continuation) of two loops to the case of framed loops.
Given two loops $\alpha^f$ and $\beta^f$ coincident only at an single
intersection point $p=\alpha^f (0) =\beta^f(0)$ we can define
the framed loop combination
$\left( \alpha \ast \beta \right)^f$ to be the framed loop which
is the ordinary product of loops with a continuous direction field,
i.e. $\theta_\alpha(1) = \theta_\beta(0)$.  This product, $\ast$, is
distinguished from the product on the abstract algebra to be defined in the
next section.

\subsection{An equivalence relation for framed loops
and the algebra ${\cal LA}^f$ }

In the usual loop representation we are not interested in
the loops themselves, but only in  equivalence classes of loops.  For
$SU(2)$ Wilson loops, these include the Mandelstam identities which
arise from the
traces of $2\times2$ matrices. We extend these relations
for framed loops.  Guided by the sloop hypothesis, we construct
an algebra of
framed loops modulo a set of equivalence relations.  This algebra we denote
${\cal LA}^f$.

Recall that given a connection the identities satisfied by Wilson
loops may be
implemented on the free vector space of formal sums of
{ \it single } loops by requiring
that if
\f
\sum_i c_i T [\alpha_i, A]=0
\nonumber
\ff
{\it for all connections}, then these loops are linearly dependent
\f
\sum_i c_i \alpha_i =0
\label{fvmandle}
\ff
on the free vector space of single loops.  (For $SU(2)$ Wilson loops, this
formulation in terms of single loops and this equivalence
relation is equivalent the loop representation with multiloops.)
When the product on the free vector space is defined as
\begin{equation}
\left(\sum_j c_j \alpha_j \right) \cdot \left( \sum_k d_k \beta_k \right) =
- \sum _{j,k} c_j d_k \left( \alpha_i \ast \beta_j + \alpha_i \ast
\beta_j^{-1} \right)
\label{uprod}
\end{equation}
these Mandelstam relations  define an
ideal, so that the quotient of the vector space by the ideal defines an
algebra, which is usually called the ``holonomy algebra'' as a reminder
that the Mandelstam relations are augmented by equivalence under holonomy.
We would like to generalize these identities for framed loops.  However, as no
notion of $q$-deformed holonomy exists (to our knowledge)
we must fall back on a purely
combinatoric definition and thus have to { \it conjecture} that the
equivalence relations defines an ideal.  Fortunately, the
hypothesis that the Kauffman bracket relations hold in the limit
of small loops suffices to defines the equivalence relations and
the resulting algebra.

We begin by defining a set of  equivalence relations on the free vector
space ${ \cal FL}^f$ of framed loops.
The first two relations are taken over from the usual loop algebra.
The first is retracing, for a single loop
\f
\alpha^f \ast \eta^f \ast \left(\eta^f\right)^{-1} = \alpha^f.
\label{retracing}
\ff
where $\eta^f $ is an arbitrary framed path of the loop $\eta^f \ast
\left( \eta^f \right)^{-1}$ beginning
at the base point of the framed loop $\alpha^f$.
The second identity results from reparametrization
invariance, for any function $f:I \rightarrow I$
\f
\gamma^f  (s) = \gamma^f (f(s)).
\label{reparam}
\ff
While ``accelerating'' the parameterization of loops has no effect on
framing, if the reparametrization does reverse the orientation of any
loop then the direction field must be reversed as well.

The remaining relations have no counterpart in ordinary loops.
One set has to do with twisting of a single loop.  It is determined
by the  sloop hypothesis to be
\f
\kd{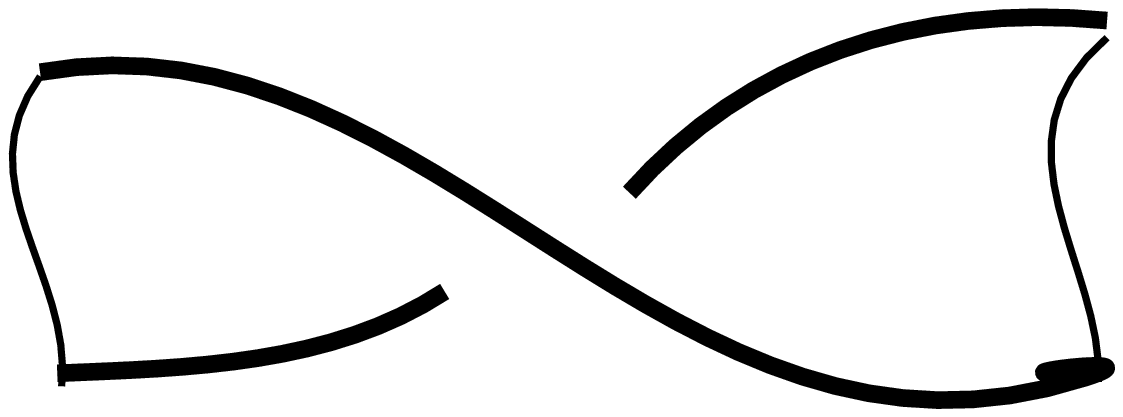} \sim \kd{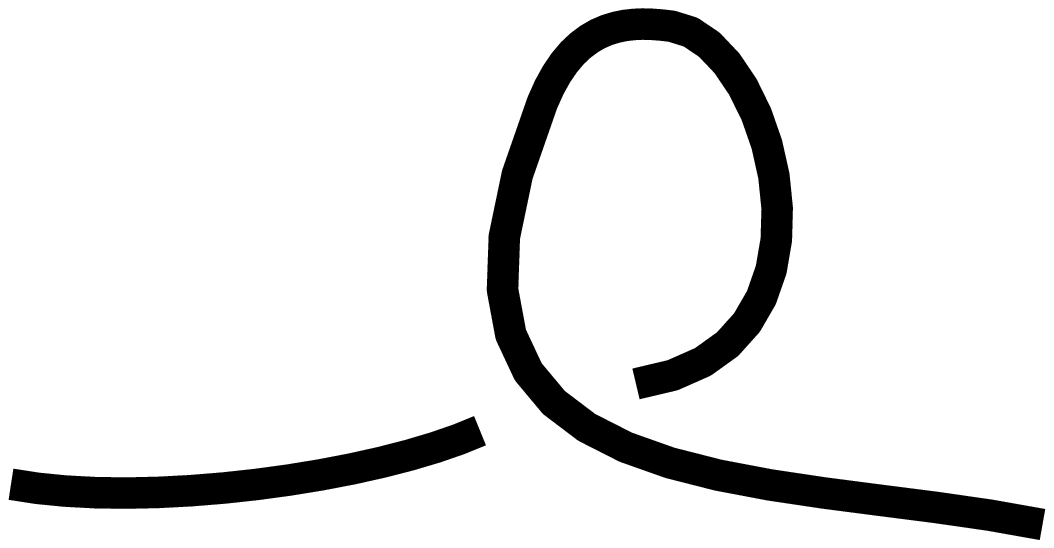} = -A^{-3} \pic{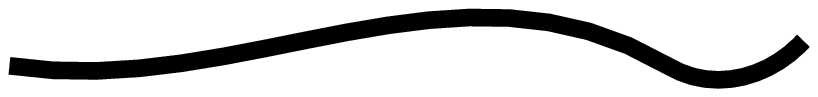}{5}{3}{0}{0}
\nonumber
\ff
In addition, the sloop hypothesis determines that the relations between
the different kinds of touching and exchanges must be given by the
skein relations,
\begin{eqnarray}
\kd{uptouch} &=& A^{-1} \kd{collision} + A \kd{wedges}
\label{upmandle}
\\
\kd{dntouch} &=& A \kd{collision} + A^{-1} \kd{wedges}
\label{dnmandle}
\end{eqnarray}
for the different
touches and reroutings at a simple intersection, $\kd{cross}$.
We may note that at $A=-1$ the framing must be
irrelevant, and hence the two independent intersections
$\kd{uptouch}$ and $\kd{dntouch}$ reduce to an ordinary
intersection.  We see that both identities reduce
to the Mandelstam identity, written in the binor notation.
Further, there is a linear combination of $\kd{uptouch}$ and
$\kd{dntouch}$ that does satisfy the ordinary Mandelstam
identity.  The coefficient $B$ of Eq. (\ref{bmandle}) can easily
be computed from Eq. (\ref{upmandle}) and Eq. (\ref{dnmandle}) to be
\f
B= {1 \over A + A^{-1}}.
\ff
Thus, these are an extended, combinatorial form of the equivalence
relation on
the free vector space of $SU(2)$ loops, Eq. (\ref{fvmandle}). All of
the relations of the holonomy algebra are included in the extended
set of equivalence relations.  This extension arises from the new
elements $\kd{uptouch}$ and $\kd{dntouch}$ which account for
framing of intersections.

We may now define the product of two equivalence classes of
framed loops.  This product,
which we will denote by $\cup$ is defined analogously to the
product of Eq. (21) on the free vector space of single loops, so that the
ordinary Mandelstam identities are satisfied by the product.
If two single, framed loops $\alpha^f$ and $\beta^f$ intersect then
$\alpha^f \cup \beta^f$ is defined to be the framed loop
in which the state at the intersection is taken to be
$\kd{cross}$ defined by Eq. (15).  This means that the
ordinary Mandelstam identities are satisfied, by $\cup$, so that
\f
\alpha^f \cup \beta^f + \alpha^f \ast \beta^f +
\alpha^f \ast (\beta^f)^{-1}=0
\nonumber
\ff
This will be sufficient to guarantee that the product $\cup$ is
associative and commutative on the equivalence classes of framed
loops.

\begin{figure}
\begin{tabular*}{\textwidth}{c@{\extracolsep{\fill}}c@{\extracolsep{\fill}}}
\pic{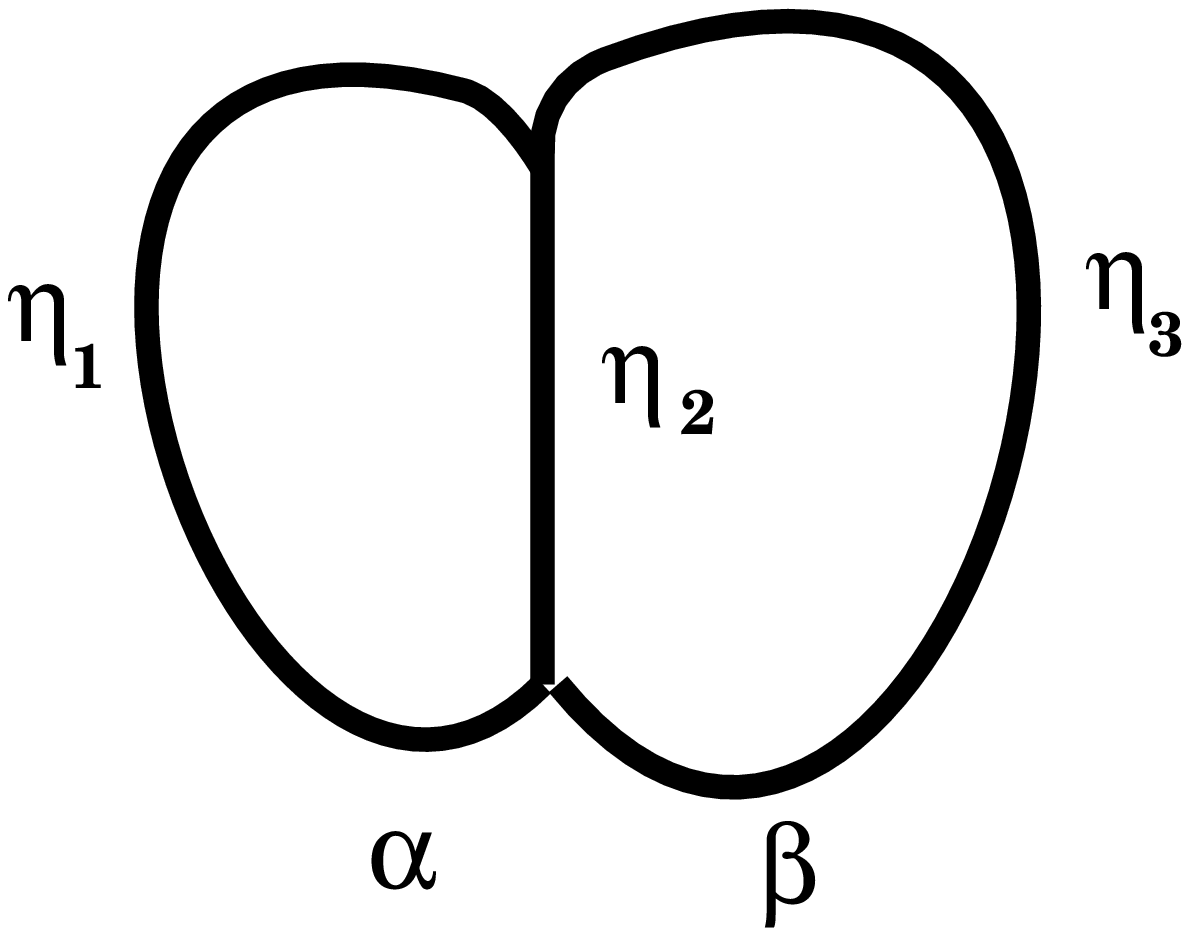}{75}{5}{0}{0}&
\pic{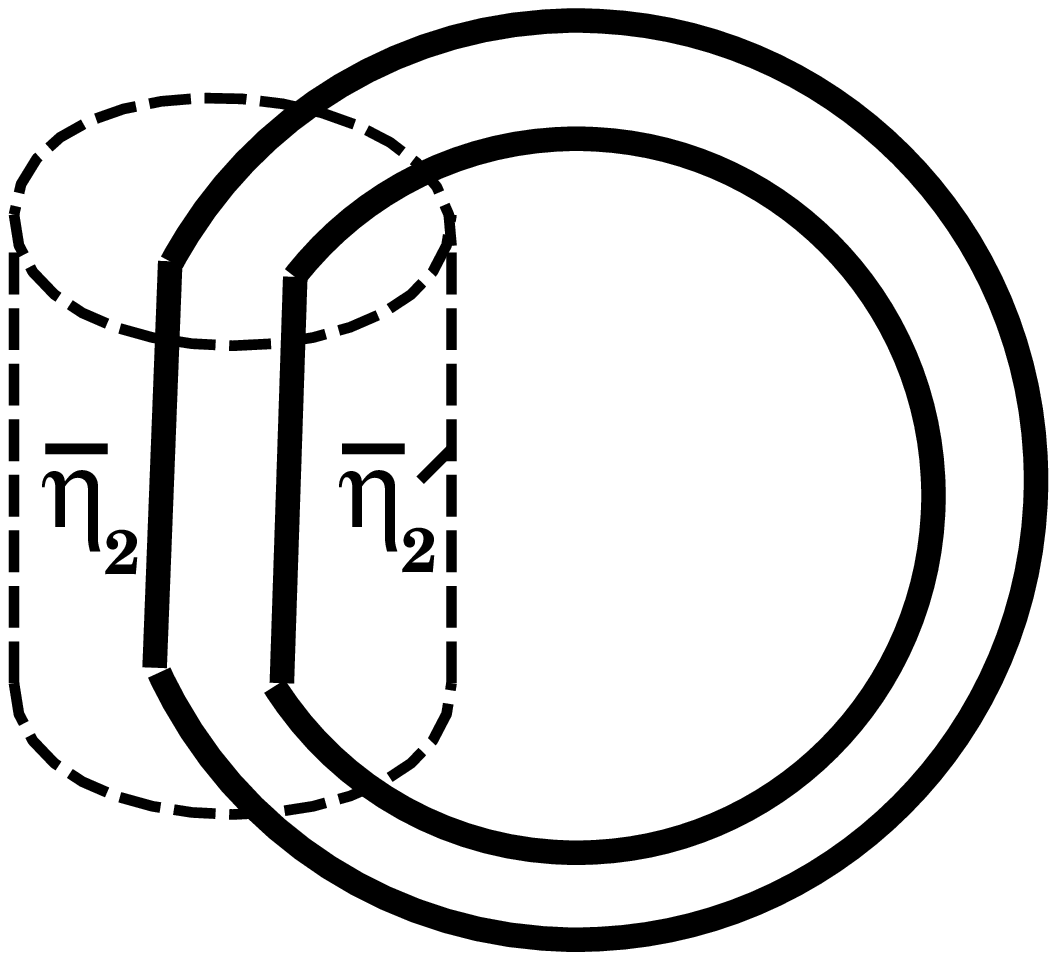}{105}{5}{0}{0} \\
a & b \\
\end{tabular*}
\caption[Framing on Retracings]{The framing on the retraced path
$\eta_2^f$ in (a.) is constructed so that the linking of the closure of
the two segments $\eta_2$ and $\eta'_2$ vanishes, i.e.
$L\left(\bar{\eta_2}, \bar{\eta'_2} \right)=0$ for the simple tangle in (b.).}
\label{retrace}
\end{figure}

Next, we define the product $\cup$ in the case that  two loops trace
a common path as in Fig. (\ref{retrace}).  Consider two framed
loops $\alpha^f=\eta_1^f \ast \eta_2^f $ and
$\beta^f = \eta_2^f \ast \eta_3^f$ constructed from the
paths $\eta_1^f, \eta_2^f $ and $\eta_3^f$ shown in Fig. (\ref{retrace}).
Note that
because of the  global nature of the framing we can take the
framing normal to the common path $\eta_2^f$.
(We can change the framing of any line
at will using the identity (24).)
There is a linking number associated with the overlap segment of two loops.
We may draw a box around the common
path such that the points $\eta_2^f(0)$ and $\eta_2^f(1)$
are embedded in the floor and ceiling
of the box (see Fig. [\ref{retrace}b]).  Closing $\eta_2$ to the tangle
$\overline{\eta_2}$,
we ask that the linking number vanishes.
The vanishing of the linking number over the
common path of $\alpha^f \cup \beta^f$ means that it can be thought
of as the limit of a sequence of unlinked paths.

The elements of the ``extended holonomy equivalence classes'' on ${\cal FL}^f$
defined by the relations Eqs. (\ref{retracing}), (\ref{reparam}),
(\ref{upmandle}), and (\ref{dnmandle}) will be denoted by $\tilde{\alpha}^f$.
The algebra constructed from these elements
$\tilde{\alpha}^f$ with the product $\cup$ is an abelian, associative
algebra, which we will call the framed loop algebra and
denote ${\cal LA}^f$.

The communtivity and associativity of
${\cal LA}^f$ follow, as in the $SU(2)$ case, directly from the Mandelstam
relations.  The key observation is that the usual equivalence
$\alpha=\alpha^{-1}$ in the usual holonomy algebra is also true on
${\cal LA}^f$, $\tilde{\alpha}^f = \left( \tilde{\alpha}^f \right)^{-1}$.
This follows both from the definition of the direction field framing of the
inverse and also from the inclusion of all the
equivalence relations for normal, unframed loops in the
extended equivalence relations.

\subsection{The operator algebra on framed multiloops}

We have defined a formal algebra, ${\cal LA}^f$.
We now construct a representation of this algebra
and use it to define the corresponding quantum theory.  This
first step to do this is to express the algebra as a formal
algebra of linear operators.  To do this we define an
operator $\hat{T}_q[\alpha]$ associated to each element
$\tilde{\alpha}^f$ of ${\cal LA}^f$.  The subscript $q$ on the operator
$\hat{T}_q[\alpha]$
means that it is associated with framed loop $\tilde{\alpha}^f$.
Thus, since the label $q$ and the superscript $f$ are redundant, we drop the
$f$.  We define the operator product so that
\f
\hat{T}_q[\alpha] \hat{T}_q[\beta] \equiv T_q[\alpha \cup \beta].
\ff
The  algebra of the operators  $T_q[\alpha ]$
is associative and commutative by virtue of the properties of  $\cup$.
This also means that the $T_q[\alpha ]$ will satisfy the ordinary
Mandelstam identities (modulo framing factors associated
with the twistings of the loops.)
However, not
all of the relations satisfied by the $\hat{T}_q[\alpha]$
agree with the relations defined
for Wilson loops of smooth $SU(2)$ connections.
In particular, in the case in which limits are taken in which
loops are shrunk down, we find a deformation of the usual
relations satisfied by $SU(2)$ holonomies.  This is forced
by the requirement that the Kauffman bracket identities
are satisfied for sloops.
To see this let $\beta^f(s,t)$ be a one parameter family
of unknots such that $\beta^f(s,0)=\beta^f(s)$ and
$\beta(s,1)=e$, the identity loop at the base point,
for all $s \in I$.  If the framing is such that
$L[\beta^f(s,t)] =0 $  and $L[\alpha^f, \beta^f(s,t) ]=0$  for all
$t \in I$ and loops $\alpha^f$ we have,
\f
\lim_{t\rightarrow 1} \hat{T}_q[\beta(s,t)] \hat{T}_q[\alpha]
=d \hat{T}_q[\alpha]
\ff
The identities given already determine
\f
d=-q -q^{-1}.
\ff
\begin{figure}
\pic{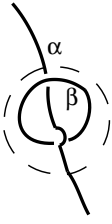}{80}{5}{20}{0}
\caption{For loops contract able to a point the linking number changes
the limit.  In this case, the limit as $\beta$ shrinks to a point
is $-A^4-A^{-4}$.}
\end{figure}
The cases in which $L[\alpha^f, \beta^f(s,t) ] \neq 0$ are also
determined by the identities, for example in the case
$L[\alpha, \beta^f(s,t) ]=1$, shown in Fig. (3) we
have\cite{KL}
\f
\lim_{t \rightarrow 1} \hat{T}_q[\beta(s,t) ]\hat{T}_q[\alpha]
= ( -A^4 -A^{-4} ) \hat{T}_q[\alpha].
\ff

\subsection{The $q$-spin net basis}

An independent basis for the algebra ${\cal LA}^f$ is given by
linear combinations of framed loops labeled
q-spin nets of $SU(2)_q$.  A q-spin net is a labelled
graph with a vertex set of arbitrary
valence.  Each edge is labeled by an
integer $j$ taken from the set $1,2,...,r-1$.
 Vertices are labeled by additional sets of integers,
describing how the singlet representation may be extracted from
the product of incident edge representations.
For each valence there must be at least one way to extract the
singlet, which leads to certain admissibility conditions.  For the
trivalent case, there is a unique way and the admissibility
conditions for $(l,m,n)$ require that $l+m-n, l+n-m$
and $ m+n-l$ are positive and even and that $l+m+n \leq 2r-4$ \cite{KL}.

Given a q-spin net we may construct a representation of ${\cal LA}^f$
by a simple prescription.  Each edge
labeled by an integer $n$ is written as a linear combination of
terms in which $n$ lines transverse the same curve, with possible
braidings.  These are given by the formula \cite{KL}
\f
\pic{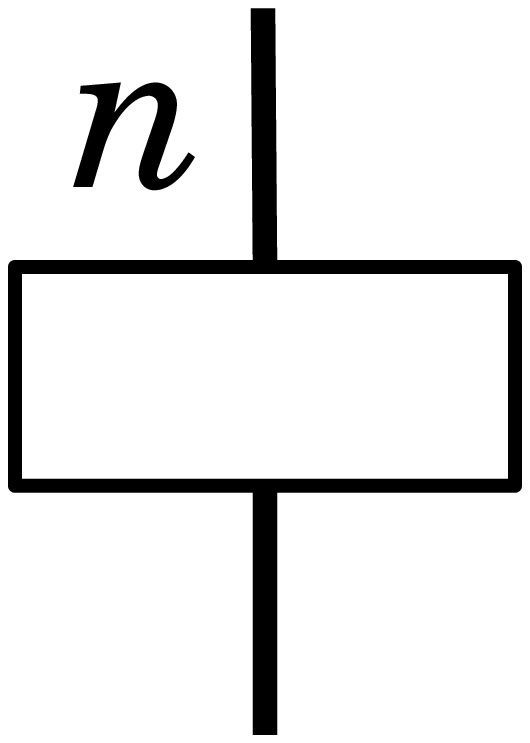}{30}{-15}{-1}{2}= { A^{2n-2} \over [n]!} \sum_{\sigma \in S_n}
\left(A^{-3}
\right)^{t(\sigma)} \pic{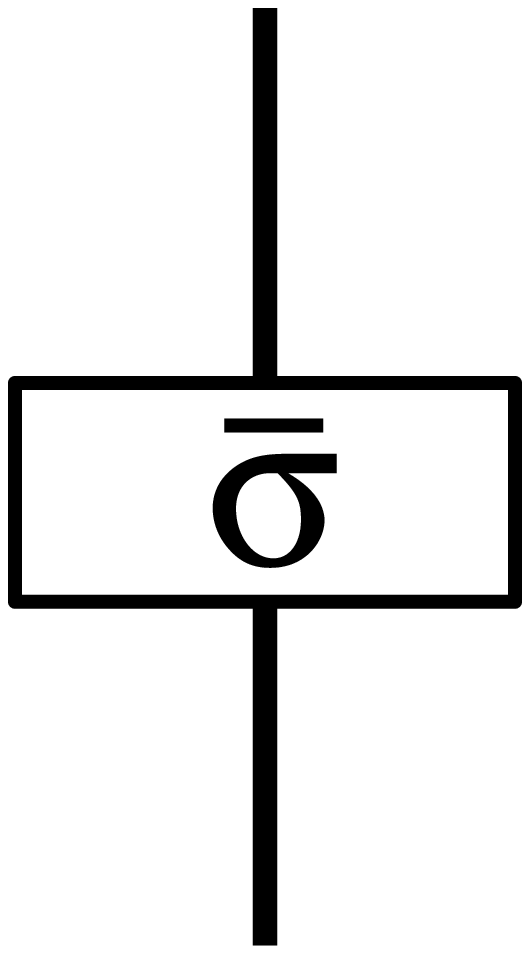}{35}{-15}{-1}{2}
\label{qsym}
\ff
where the ``quantum integer'' $[ n ]$ is defined by
\f
[ n ] = {q^{n} -q^{-n} \over q-q^{-1}}, \nonumber
\ff
the factorial is defined as $[n]!=[n][n-1] \ldots$,
and $\sigma$ is an element of the permutation $S_n$ with a minimal braid
representation $\overline{\sigma}$ consisting of the minimum number of
over crossing elements $\sigma$ of the braid group.  For instance, for $n=2$
we have
\begin{eqnarray}
\kd{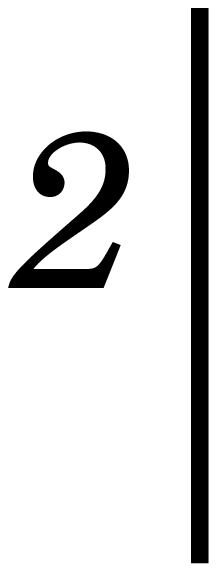} & = & { A^{2} \over [2]!} \left( \kd{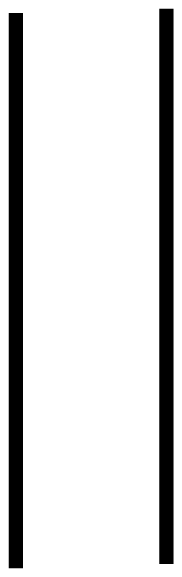} + A^{-3}
\kd{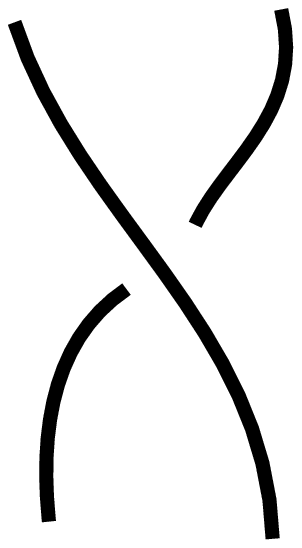} \right) \nonumber \\
& = & \kd{2para} - {1 \over d} \kd{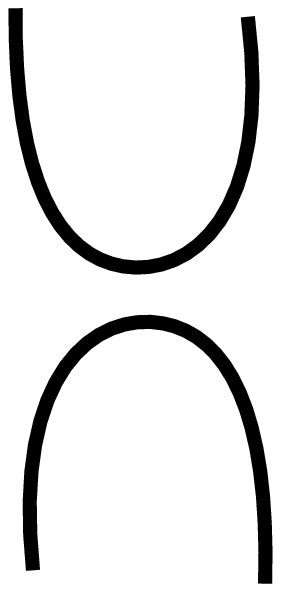}
\end{eqnarray}
in which the over crossing on the first line represents the
``touching from the top,'' $\kd{uptouch}$.
We may note that these satisfy some simple identities \cite{KL},
\begin{eqnarray}
\pic{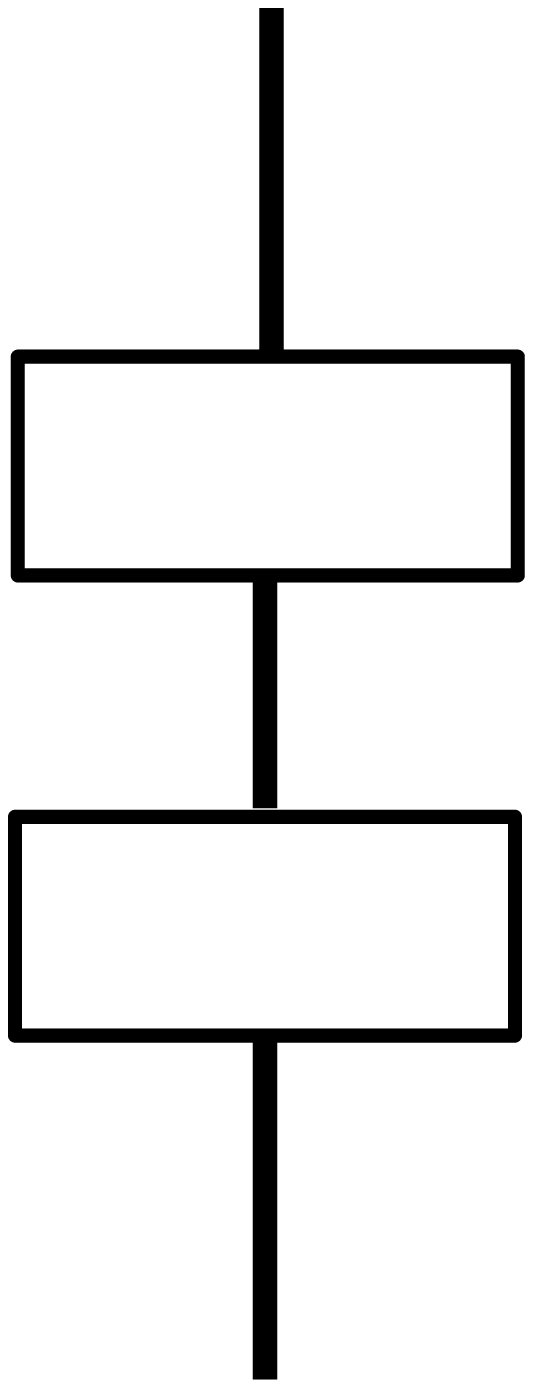}{50}{-23}{0}{0} & =& \pic{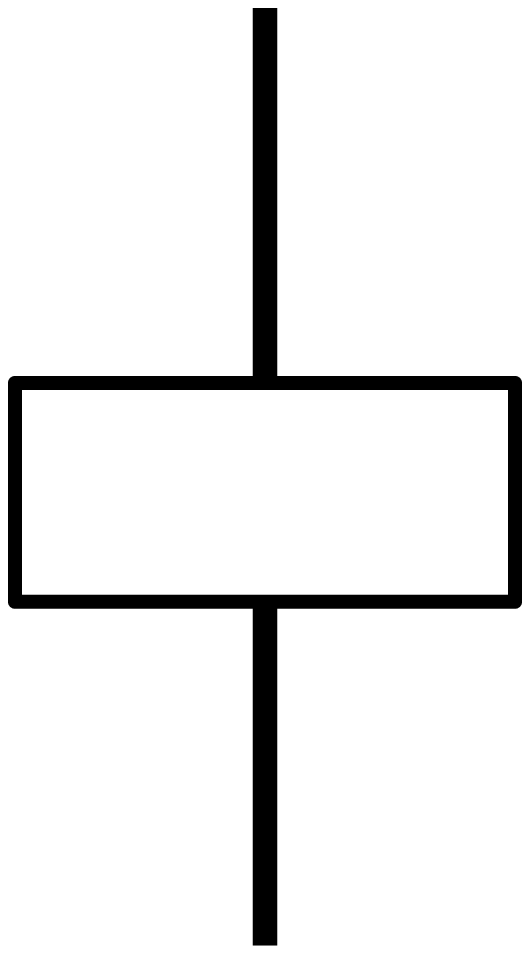}{30}{-10}{0}{0}
\nonumber \\
\pic{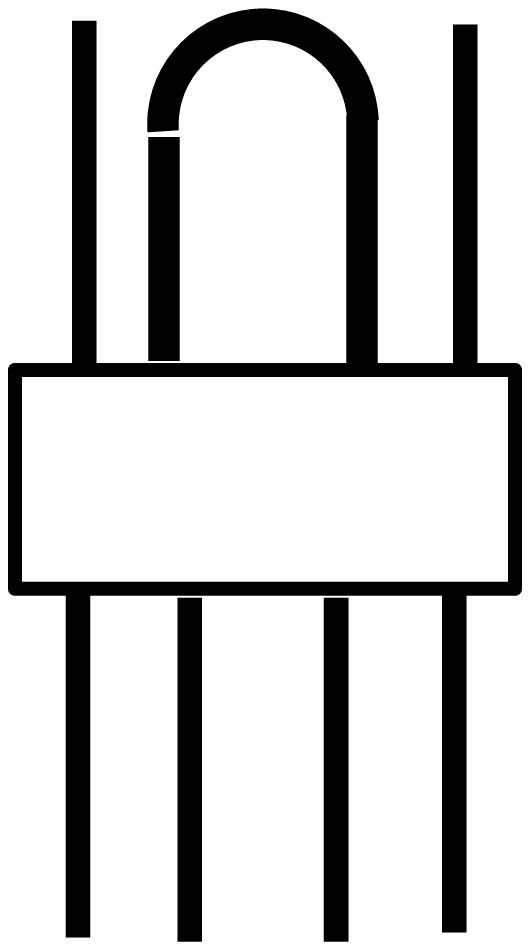}{40}{-15}{0}{0} & = & 0 \nonumber \\
\pic{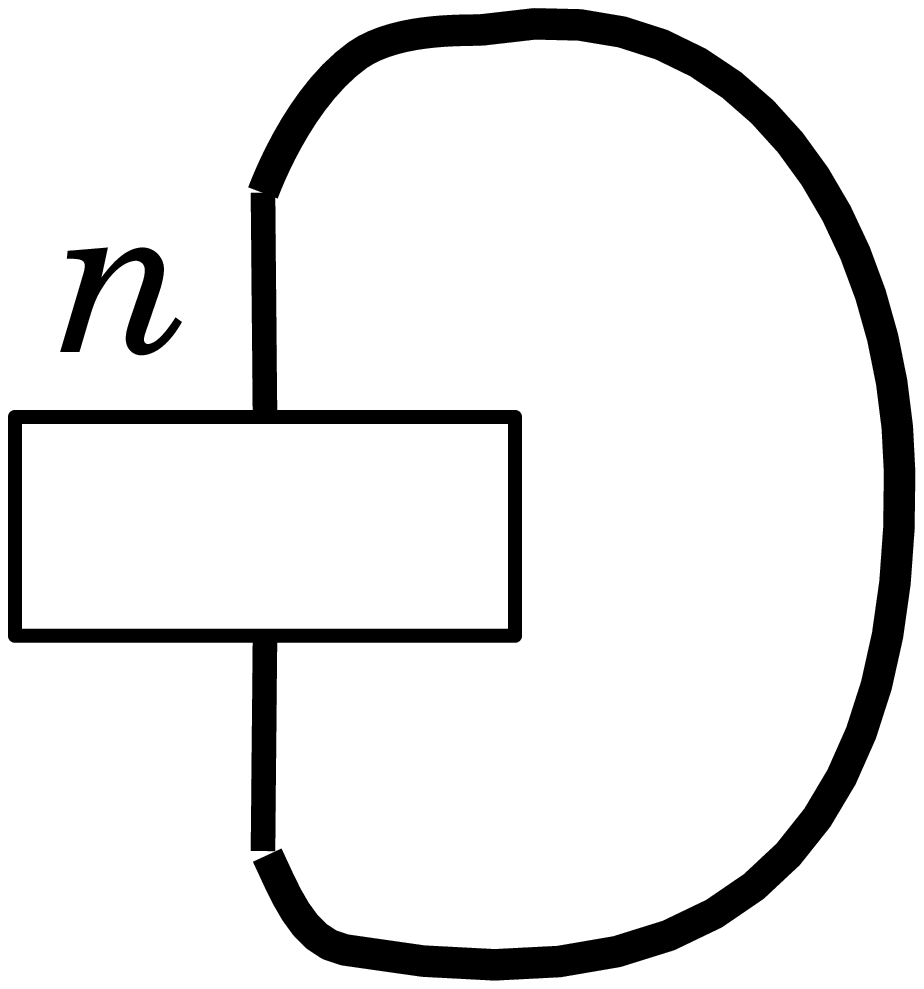}{40}{-15}{0}{0} & = &(-1)^n [n+1]
\label{closedloops} \\
\pic{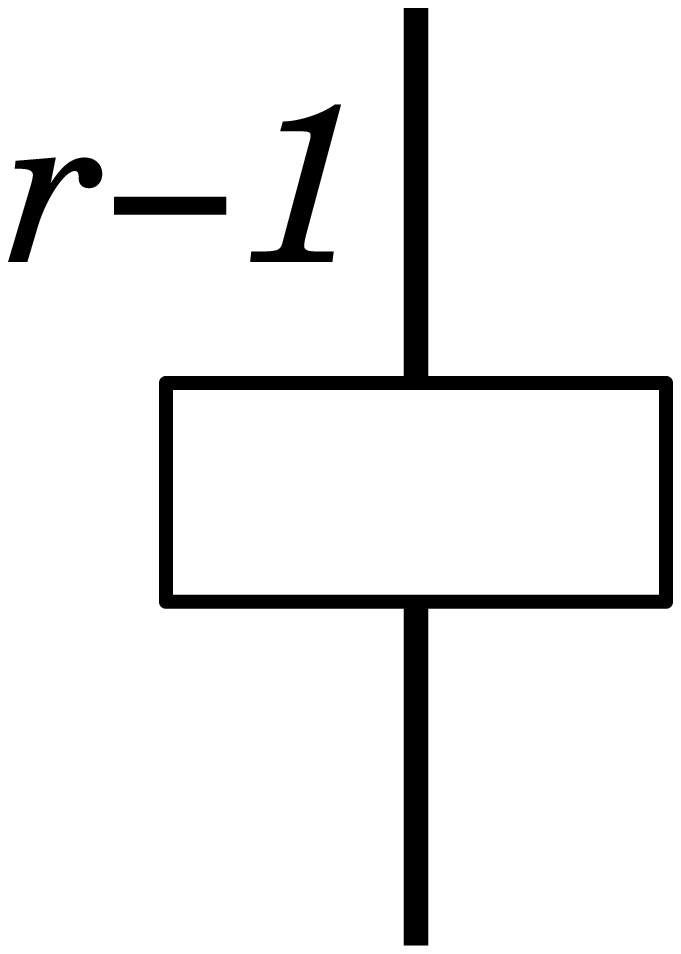}{40}{-15}{0}{0} & = & 0.
\end{eqnarray}
\begin{figure}
\begin{tabular*}{\textwidth}{c@{\extracolsep{\fill}}c@{\extracolsep{\fill}}}
\pic{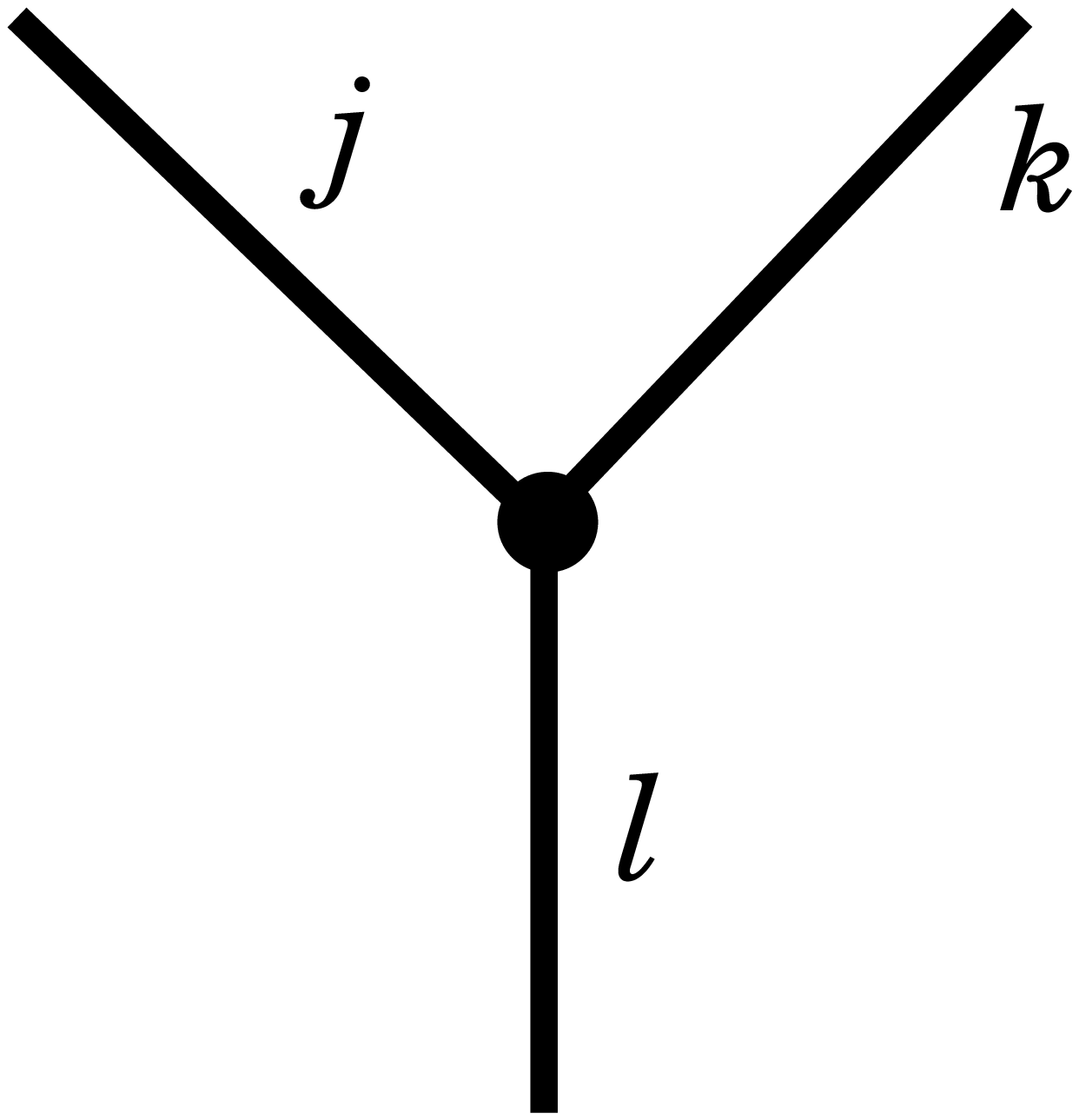}{70}{-5}{20}{0} & \pic{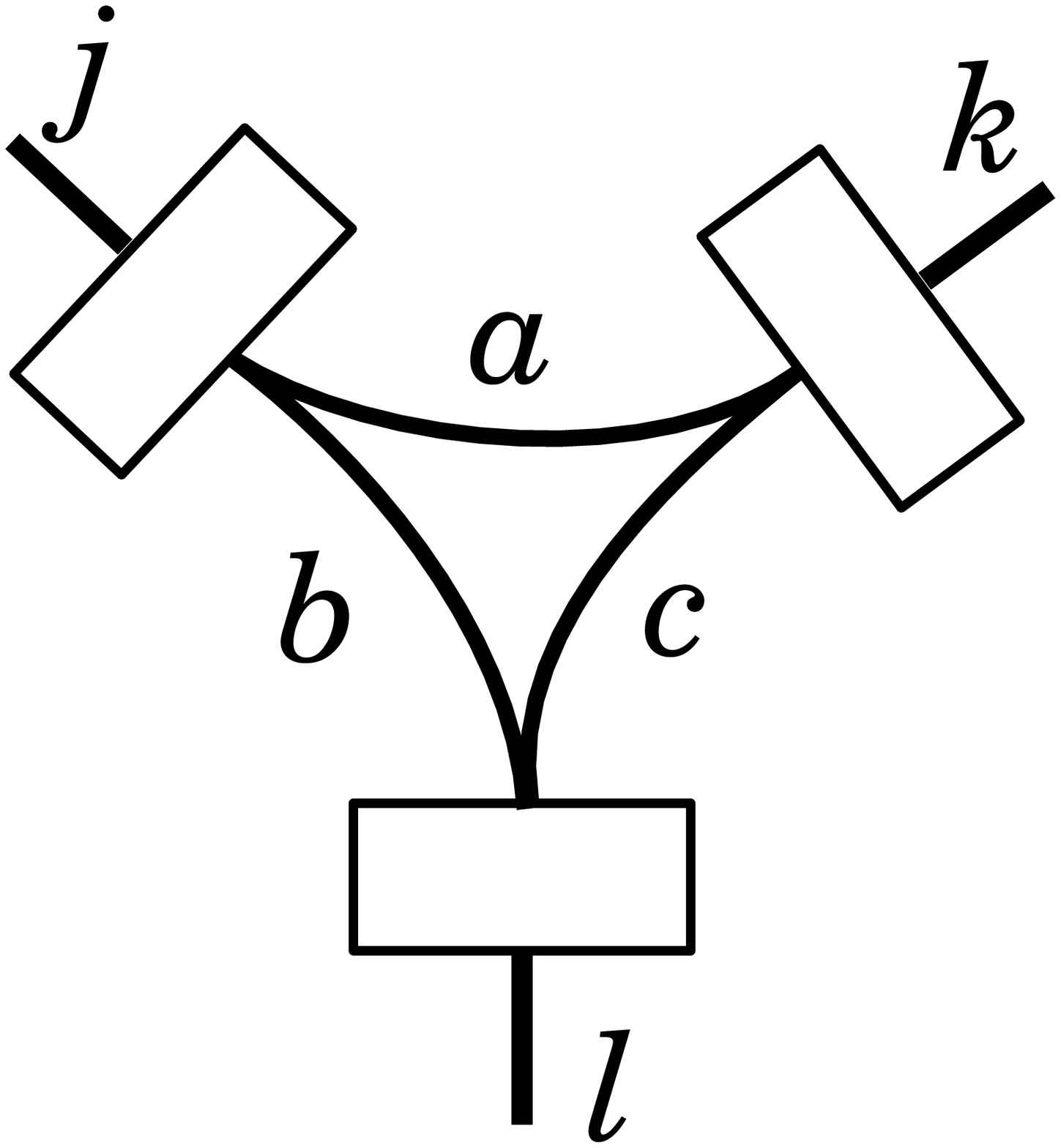}{80}{-5}{0}{20} \\
a & b \\
\end{tabular*}
\caption{The trivalent vertex (a.) is decomposed into three projectors
as in (b.) with $a = (j+k-l)/2$, $b=(k+l -j)/2$, and $c=(j+l-k)/2$.}
\label{trivalent}
\end{figure}
Trivalent intersections are decomposed according
to Fig. (\ref{trivalent}).
A vertex of higher valence requires an additional label because
there is more than one way to combine the $SU(2)_q$
representations of its incident edges into an $SU(2)_q$ singlet.
Thus there is a finite dimensional linear space
to each $n$ valent vertex ($n>3$) with
incident edges labeled by the $j_i$.
A basis for these vertices may be constructed in the following way.
One first picks an arbitrary ordering of the edges
which are incident on
the vertex.
One then decomposes
the $n$ valent vertex into a
combination of trivalent vertices as illustrated in Fig. (\ref{nvertexfig}).
The number of internal vertices is $l=1+(n-4)=n-3$. A set of
linearly independent states associated with the $n$ valent vertex are
label sets ${l}$ of the ordering of external lines and the
internal representations $i_1, i_2, \ldots, r-1$ on
the internal lines so that the
trivalent vertices created by this procedure are admissible.

\begin{figure}
\pic{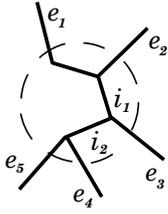}{80}{-15}{20}{0}
\caption{The decomposition of a higher valent intersection into trivalent
intersections at a point. The first two incident edges are joined
to a new internal edge $i_1$ at the
first vertex.  Then $i_1$ and $e_3$  are joined into a trivalent
vertex with a new internal line $i_2$.  The process continues until
there are two external vertices left which are
joined into the last three vertex with the last internal line, in this
case $i_2$.}
\label{nvertexfig}
\end{figure}
Three comments should be made about this labeling.  First internal
edges have zero length in the manifold $\Sigma$, so that all the trivalent
vertices in this ``blowing up at the vertex" are at the same point of
$\Sigma$ as the original $n$-valent vertex.  Second, given a different
labeling of the external edges, the same procedure will yield a different,
orthogonal basis.  Each relabeling
of the edges of the graph thus is represented by a unitary transformation
in each of the spaces associated with the
vertices.  Finally, a decomposition of the vertices of a spin
network may be given by arbitrarily labeling all of its edges,
which induces a labeling of the edges of each vertex.

Given a q-spin net $\Gamma^q$, we have, after expanding the terms,
an element of ${ \cal LA}^f$.
Conversely, it is straightforward to show that given any
framed multiloop $\gamma^f$ we can construct a unique formal
linear combination of quantum
spin networks $\Gamma^q_i$
so that
\f
 \gamma^f = \sum_i  c_i \Gamma^q_i
\ff
The construction follows an algorithmic procedure, which
extends (because of the extension of the Temperly-Lieb
algebra to finite loop segments representing holonomies)
the algorithm of Kauffman and Lins \cite{KL}.
We proceed by labeling abstract edges by framed loops.  Each edge
may carry a number of segments representing framed loops with common
support between vertices.  Vertices occurs where the support of
the loops changes.  Given a labeling of the edges the framing dependence
of the loops may be expanded as a sum of $q$-symmetrized lines,
defined by Eq. (\ref{qsym}).
Independent routings involving $n$
segments transversing a single edge are elements
of the ``extended'' (in the sense of holonomy) Temperly-Lieb
algebra ${\cal T}_n$.  But,  as described in
Kauffman and Lins, a basis for ${\cal T}_n$ is described in terms of
projection operators and retracing elements such as $\kd{wedges}$ which,
because of equivalence under retracings, pull back to the
adjacent vertex.  The result is an
expansion of the framed multiloop as a sum of terms of $q$-symmetrized
lines incident on a vertex - a $q$-spin network.

We have sketched a demonstration that the q-spin nets provide a
representation of the framed loop algebra ${ \cal LA}^f$.
It remains to show that the q-spin nets
are independent under the identities of Section 2.4.   The
demonstration will not be given in full here,
but we note that it is an extension of
Proposition 2 of Kauffman and Lins.  The basic step uses the fact, already
mentioned, that to each edge with $n$ common segments
we may associate elements
of the Temperly-Lieb algebra ${\cal T}_n$ such that the different
$q$-spins label orthogonal projection operators.

Finally, we note again that in the case of valences $n>3$ the
uniqueness of the $q$-spin network basis is only up to arbitrary
relabeling of the edges of the graph, as different labelings induce
unitary changes of basis at each vertex of valence $n>3$.

\section{The framed loop representation
in the $q$-spin net basis}

Define ${\cal H}^q$ to be the space of
functionals on ${\cal LA}^f$.  Introducing ``bra" states
$\bra{\alpha^f}$ for $\tilde{\alpha}^f \in
{\cal LA}^f$ this may be written as,
\f
\Psi [\alpha^f ] = \braket{\alpha^f} {\Psi } .
\nonumber
\ff
On this space of states we may define a representation of the
framed commutative loop algebra by
\f
\bra{\alpha^f} \hat{T}_q[\beta] = \bra{\alpha^f \cup \beta^f}
\nonumber
\ff
By the product properties (30) this defines
a faithful representation of the algebra.
One of the key results of the loop representation is the existence of
the spin network basis\cite{spinnetcl}\footnote{This result has been made
mathematically rigorous in the context of diffeomorphism invariant
measures on the connection representation by
Baez \cite{baez-spinnet}.}.  We showed
above an analogous result, which is that the algebra  ${\cal LA}^f$,
has an independent basis given by the q-spin nets. We may
now apply this directly to show that ${\cal H}^q$ has
an independent basis given by the q-spin nets.

Given the decomposition of a q-spin net
$\Gamma^q= \sum_i c_i \gamma_i^f$ in terms of framed multiloops
$\gamma_i^f$ we define
\f
\bra{\Gamma^q} = \sum_i c_i \bra{\gamma^f_i}
\nonumber
\ff
It follows from the independence of the $\Gamma^q$ in ${\cal LA}^f$
that these states are independent and thus provide a basis
(again up to unitary transformations at each higher than trivalent
node induced by relabeling the edges.)

We may now define the action of the $\hat{T}_q[\alpha ]$
directly on the $q$-spin net basis.  Given a $q$-spin net
$\Gamma^q$ and a loop $\tilde{\alpha}^f$, we can define a unique decomposition
of the framed loop product, of a framed loop and
a spin network
\f
\Gamma^q \cup \tilde{\alpha}^f = \sum_i c_i [\Gamma \cup \alpha ]^q_i
\nonumber
\ff
where the $[\Gamma \cup \alpha ]^q_i$ are the spin networks
produced by iterating the edge addition identity
\begin{equation}
\pic{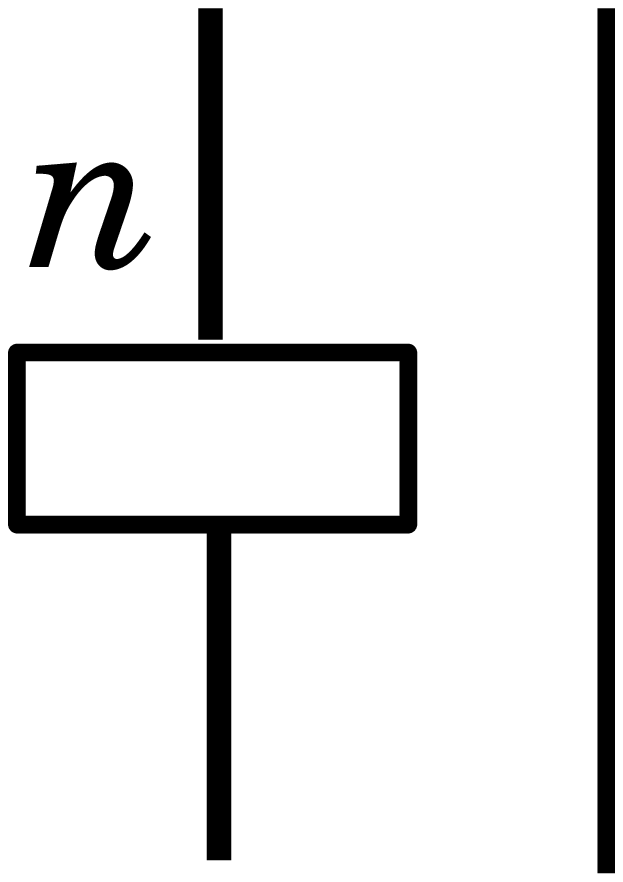}{35}{-15}{-1}{2} =\pic{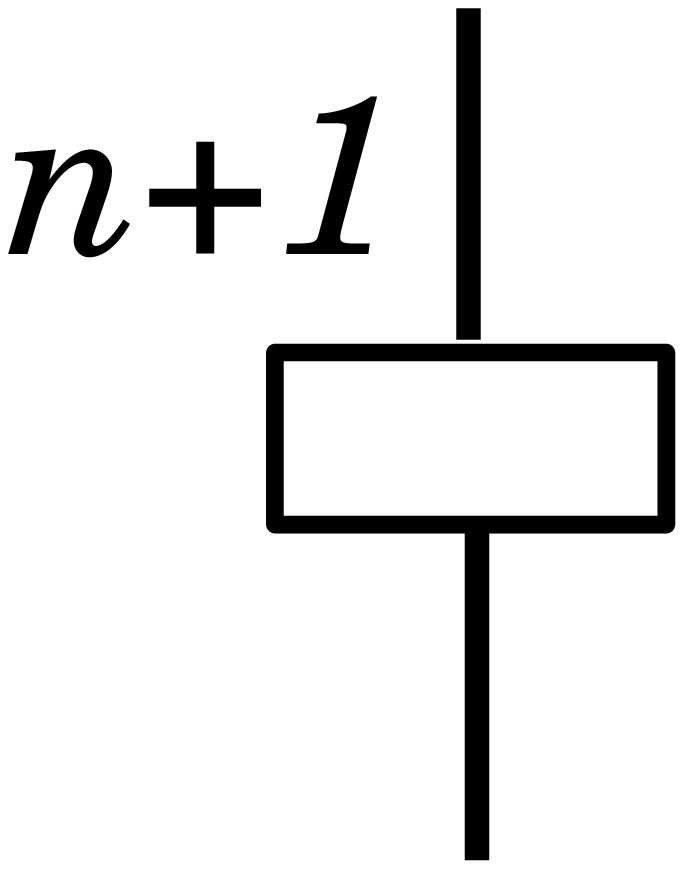}{35}{-15}{-1}{2} -
{ [n] \over [n+1]} \pic{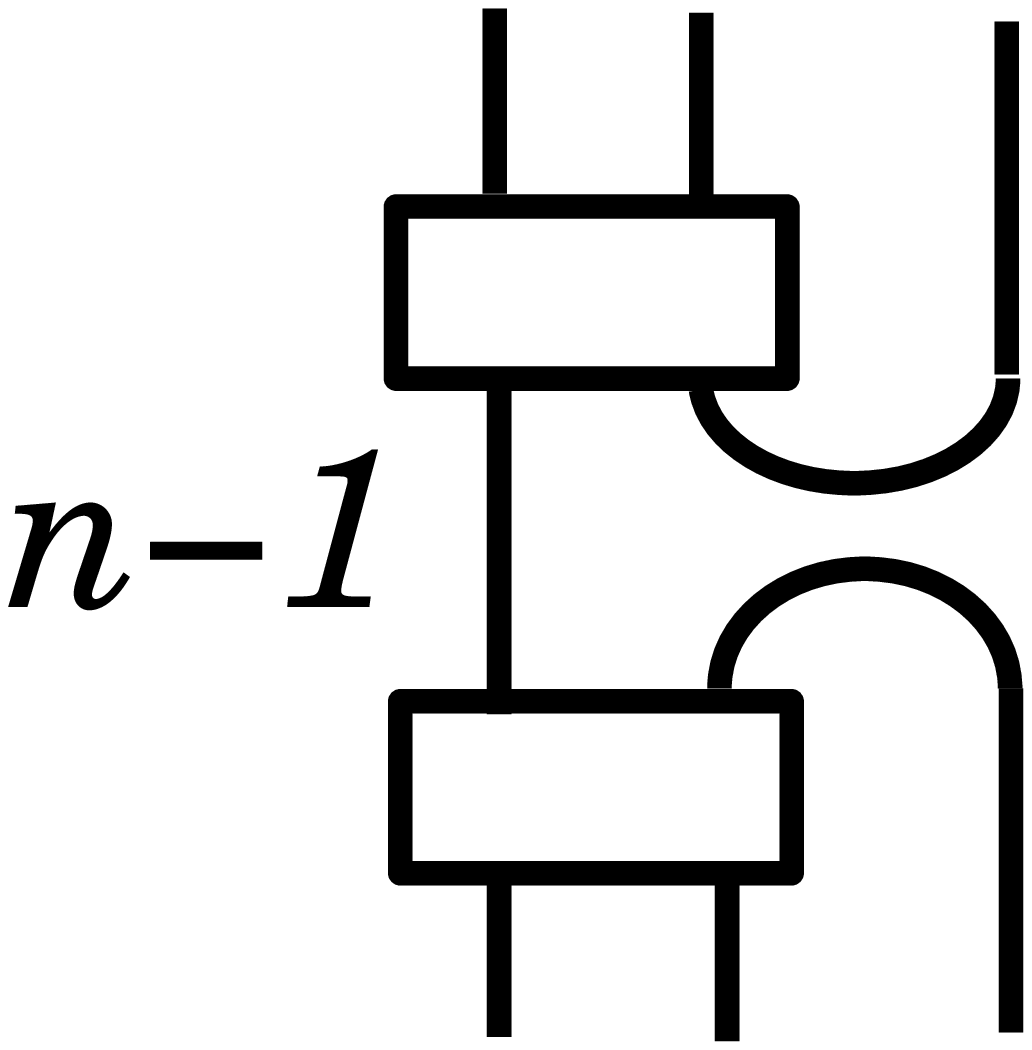}{35}{-15}{-1}{2}.
\label{edgeaddition}
\end{equation}
When adding two edges labeled by one (frequently used in the
action of the $T$ operators) this edge addition identity is simply
\begin{equation}
\kd{2para} = \kd{2line} + { 1 \over d} \kd{2cups}.
\end{equation}
Note that $\tilde{\alpha}^f$ may intersect $\Gamma^q$
either in isolated points or
in common edges.  We use the edge addition identity
in Eq. (\ref{edgeaddition}) along every
edge on which $\tilde{\alpha}^f$ and $\Gamma^q$ overlap.  If we label
the remaining edges of $\tilde{\alpha}^f$ which have no common
segment with $\Gamma^q$ with one then we have a sum of graphs
with labeled edges.  The isolated
intersection points of $\tilde{\alpha}^f$ and $\Gamma^q $ are decomposed as
$n$-valent vertices.  When they have a transverse intersection
we have a new 4-valent vertex with internal edges,
$1,1,j,j$.  When the loop and network part at a $n$-valent vertex
then we have an $n+2$ valent vertex.
We apply the same technique to each beginning or
end of an overlapped edge, which either is a new vertex or
is a change in vertex labeling when the overlapped edge
ends on an existing vertex. We may then conclude that
\f
\bra{\Gamma^q} \hat{T}_q[\alpha ] = \sum_i c_i \bra{ [\Gamma \cup
\alpha^f ]^q_i}
\ff
This gives us the action of the loop operator directly in the
spin network basis.  We may note that this formula applies at
all $A$ including $A=-1$ that corresponds to the classical case.
Thus it applies equally well to ordinary spin network states.

Finally, following the usual procedure for spin network
states \cite{spinnetcl}
we can impose an inner product on ${\cal H}^q$ extending the inner product
on spin networks to q-spin nets.
\f
< S^q | \Gamma^q > =
\delta_{S^q \Gamma^q}.
\nonumber
\ff

\section{The quantum deformed $\hat{T}_q^a$ operators}

To complete the definition of the deformed loop algebra we need
to give a definition of the ``$T^1$" operators acting on
${\cal H}^q$ and show that the result is a closed algebra.
First, we work out the action
of ordinary $\hat{T}^a[\alpha ](s)$ operators on the spin
network basis.  By extending this formula, the action of a $q$-deformed
$\hat{T}_q^a[\alpha](s)$ is defined for $q$ at a root of
unity.  We then check that $\hat{T}_q[\alpha]$ and $\hat{T}_
q^a[\beta](s)$  form a closed algebra.

\begin{figure}
\begin{tabular*}{\textwidth}{c@{\extracolsep{\fill}}c@{\extracolsep{\fill}}}
\pic{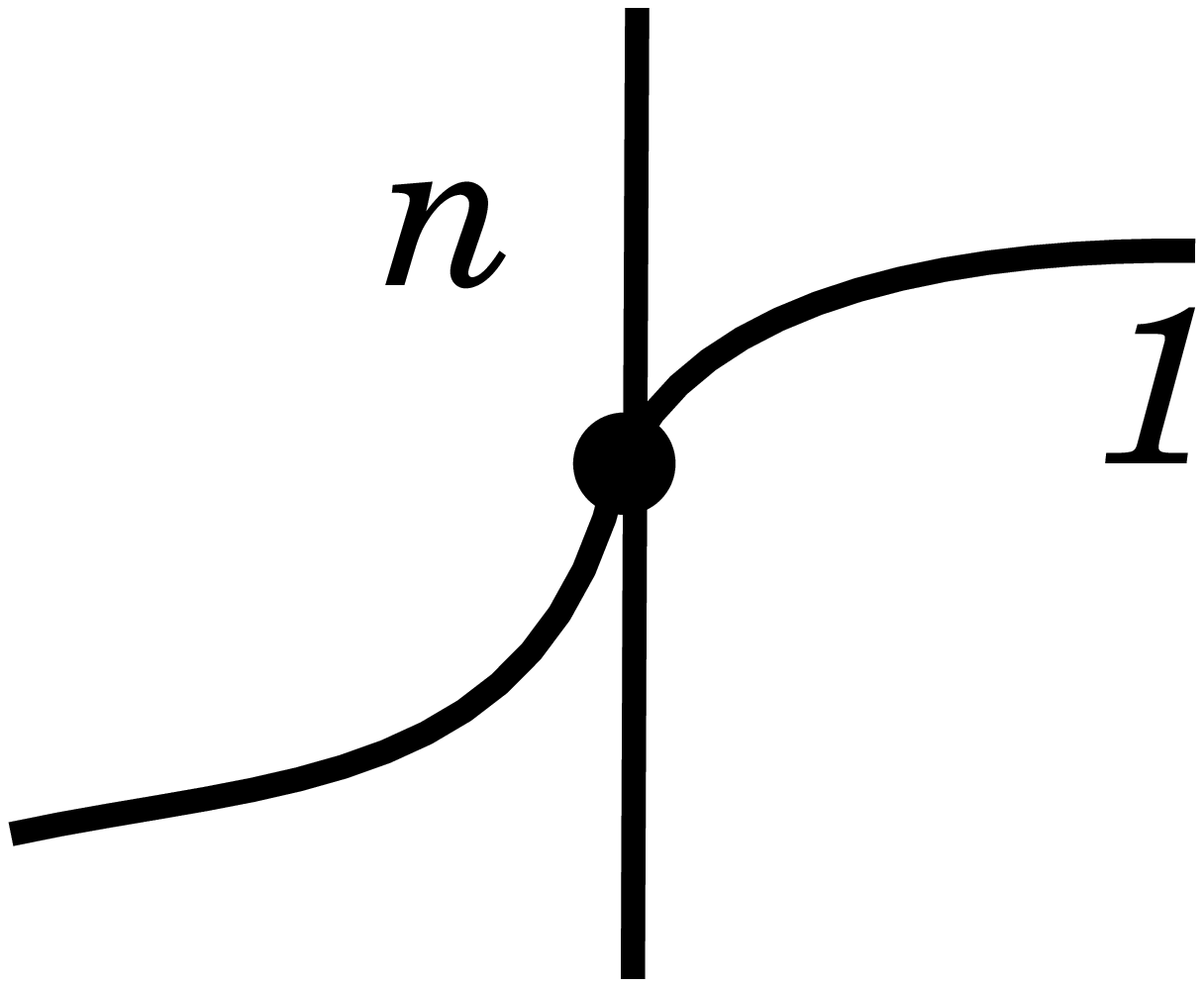}{50}{5}{20}{0} & \pic{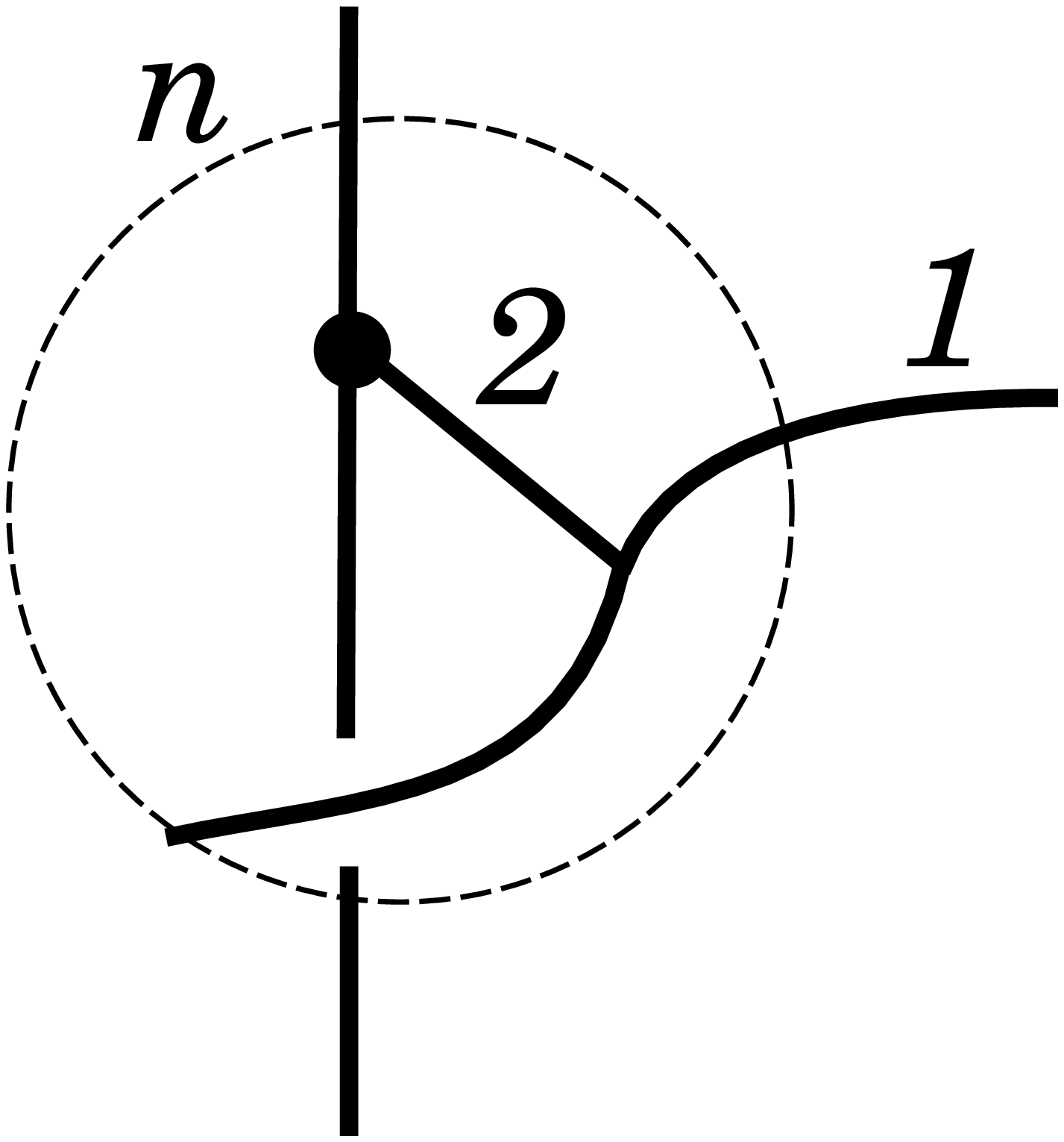}{75}{5}{0}{20} \\
a & b \\
\end{tabular*}
\caption{The action of the $T^a[\alpha]$ on a spin network}
\label{actionT1}
\end{figure}
The action of an operator $\hat{T}^a[\alpha ](s)$ on a spin network
state $\bra{\Gamma}$ is illustrated in Fig. (\ref{actionT1}).
 When the
hand at the point $\alpha (s)$ coincides with a point
on an edge of $\Gamma$ with spin $n$ a new four valent vertex
is created with incoming edges $(n,n,1,1)$ as shown.  The particular
vertex is shown in Fig. (\ref{actionT1}), it may be decomposed into
a trivalent $(n,n,2)$ vertex connected through an ``internal" $2$
line to a $(2,1,1)$ vertex.  The result is multiplied by a factor
of $n l_{Pl}^2$ and $\Delta^a [\Gamma^q , \alpha ] (s)$, the
distributional factor
\f
\Delta^a [\Gamma^q , \alpha ] (s) = \sum_I \int dt \delta^3
(e_I (t) , \alpha (s)) \dot{e}_I^a (s)
\ff
where the sum is over the edges, $e_I $ of the network.

On q-spin nets we will then {\it define} the
operator $\hat{T}_q^a[\alpha ](s)$ to act by exactly the
same diagram Fig. (\ref{actionT1}), where the vertex is
now a q-spin net vertex.  The factors we multiply
by of Eq. (47) and $j l_{Pl}^2$ are taken to be the same.
There is an ambiguity in the
definition of the action of a hand
that is not present in the ordinary $A=-1$ case.
This arises when the four valent intersection is defined in terms of
trivalent vertices according to Fig. (\ref{actionT1}b). There may be a phase
factor depending on whether the one line crosses over or
under the $n$ line.  When the vertex is created by the action of an
operator, the ambiguity is resolved by careful definition
of the regularized operator.  We will see below that there is a natural
choice in the case of the area operator.

We may now show that the quantum deformed loop operators
$\hat{T}_q[\alpha ]$ and $\hat{T}_q^a [\alpha ] (s) $ define
 a closed algebra.  We first note that the $\hat{T}_q[\alpha ]$'s
commute, by Eq. (28).
As in the case of the ordinary loop algebra, it is straightforward
to compute the commutator and verify
\f
\left [ \hat{T}_q[\alpha ] , \hat{T}_q^a [\beta ] (s)  \right ] =
j l_{Pl}^2 \Delta^a [\alpha , \beta  ](s) \hat{T}[\alpha \#_s \beta ]
\ff
where $\alpha$ is an arbitrary q-spin net, $\beta^f$
is a single framed loop and the combination
$\alpha \#_s \beta$ is a q-spin net constructed according to the following
prescription:  Break the loop
$\beta^f$ at $s$ and break the edge of the spin network $\alpha$
at the point $p=\beta (s)$.  Let the valence of the edge which
coincides with $p$ be $n$. We then  reconnect the lines with a four
valent vertex with a line labeled by two connecting the
broken $n$ line and the broken $1$ line as shown in Fig. (\ref{actionT1}).
In the case that more than one edge of the spin network $\alpha$
coincides with the point $p$, the result of the commutator is the
sum of the actions on each edge.  We may note that given the definition
of the operator on a $n=1$
line the action for arbitrary $n$ can also be recovered from the
algebra.  Also,
 as in the original case, we could express this
in terms of the strip-loop algebra \cite{ls-review}, so that the coefficients
of the algebra are non-singular.

It remains to verify the commutation relations of the
$\hat{T}_q^a[\beta ](s)$.
To do this it is helpful to define a more general
notion of these operators.  Let $\beta$ now be a general q-spin
net, and let $t$ be any parameterization of the edges.
(For example, all the edges may be ordered and the
$i$'th each may be
parameterized by $s$ running between $i-1$ and $i$.)  We
then can define an operator $T^a_q[\beta ](s)$ for every
$s$ such that $\beta (s)$ is on a single (or a ``one") line.  The
definition in this case is taken from Eq. (48)
\f
\bra{\alpha^f} \hat{T}_q^a [\beta ](s) = \sum_I j_I
\Delta^a [\alpha , \beta  ](s)
\bra{\alpha \#_I \beta^f},
\ff
where $I$ labels the intersection points where the action
is non-vanishing.

It is then straightforward to verify by direct computation
that
\f
\left [
\hat{T}^a_q[\alpha ](s) , \hat{T}^b_q[\beta ](t)
\right ] =  l_{Pl}^2 \left (    j \Delta^b [\alpha , \beta  ](t) \hat{T}^a_q
[\alpha \#_t \beta ] (s) -  j^\prime
 \Delta^a [\beta , \alpha  ](s) \hat{T}^b_q
[\beta \#_s \alpha ] (t)
\right )
\ff
where $j$ and $j^\prime$ are the $q$-spin of the lines on which
the ``hands" act in each case.
In verifying these relations, it is convenient
to use the various identities which organize the results
into the spin network basis only in the last steps of calculation.

\section{Higher loop operators and the $q$-deformed area
operator}

Once we have defined the action of the first two loop
operators the definition may be extended for loop operators
with any number of ``hands" that correspond
to points of insertion of the conjugate electric field
$\tilde{E}^{ai}$ in the classical loop algebra.  We
do this by requiring that each hand acts according to the
usual definitions of loop operators, but where the combinatorics
and framing at each hand is given by Fig. (\ref{actionT1}).
For example, the
action of $\hat{T}_q^{ab}[\gamma ] (s,t)$, for
$\gamma$ a q-spin net with $\gamma (s)$ and
$\gamma (t)$ on one segments, may be defined as
\f
\bra{\alpha^f} \hat{T}_q^{ab}[\gamma ] (s,t) = \sum_{IJ}j_I j_J
l_{Pl}^4 \Delta^{a} [\alpha_I , \gamma ](s)
\Delta^{b} [\alpha_J , \gamma ](s)
\bra{\alpha^f \#_s \#_t \gamma^f}
\ff
where $\alpha_I$ are edges of the q-spin net
and $\alpha^f \#_s \#_t \gamma^f $ is constructed by implementing
the action described above at the coincident points of $\alpha^f$,
$\gamma^f (s)$, and $\gamma^f(t)$.   There are, however, framing choices
for the operators.  Defined as the limit of a sequence of loops, these
operators have the same ambiguity that arises for intersecting framed
loops.  As an example, we give a definition of the area operator.

A $q$-area operator,
$A^q[{\cal S}]$, which measures the area of a surface
$\cal S$, is constructed
by the procedure described
in \cite{volumecl}, where the
operator we have just defined replaces the usual
$\hat{T}^{ab}[\gamma ](s,t)$.  One discovers as before that the
simultaneous eigenstates of all these operators are given by
the $q$-spin networks, or by linear combinations
of them involving the different routings in higher valent
vertices.  Following the same reasoning as in \cite{volumec1} one sees directly
that the spectrum is discrete.
The eigenvalues of  $A^q[{\cal S}]$ may be calculated  in a similar
manner as before.  Writing the operator as
\begin{equation}
\hat{A}^q \ket{\Gamma^q \, n} = \sum_I \sqrt{ \left| {1 \over 8} c_I(n)
\right| } \ket{\Gamma^q \, n}
\end{equation}
were the sum is over intersections of the surface ${\cal S}$ and
the q-spin net.  When the
q-spin net $\Gamma^q$ intersects $\cal S$ at a single
edge, the combinatorial operator $\hat{A}^q$ ``grasps'' this edge
\begin{eqnarray}
8 \left(\hat{A}^q\right)^2 \ket{\Gamma^q \, n}
&=& n^2 \pic{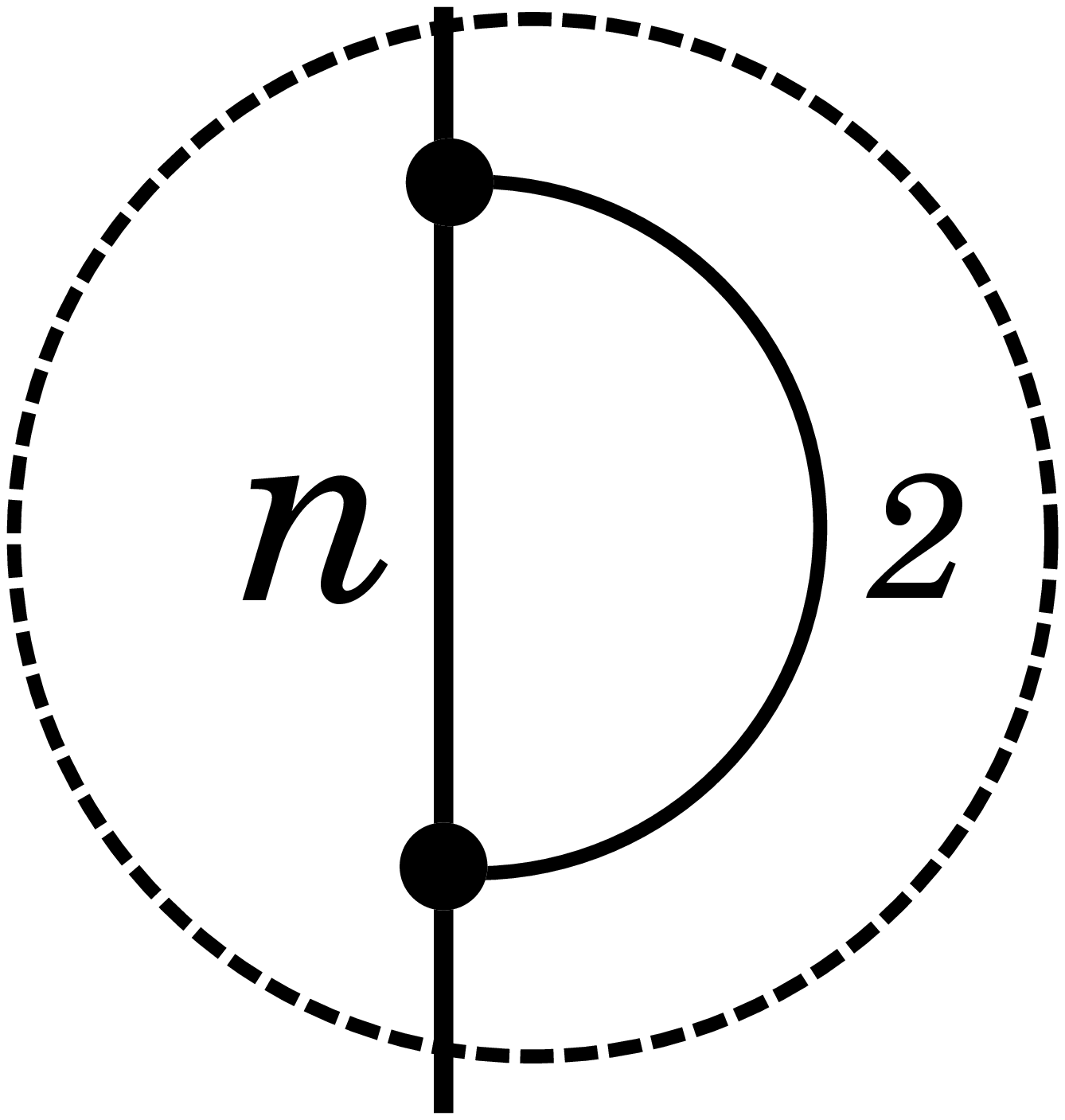}{45}{-15}{0}{0} \nonumber \\
&=& n^2 \pic{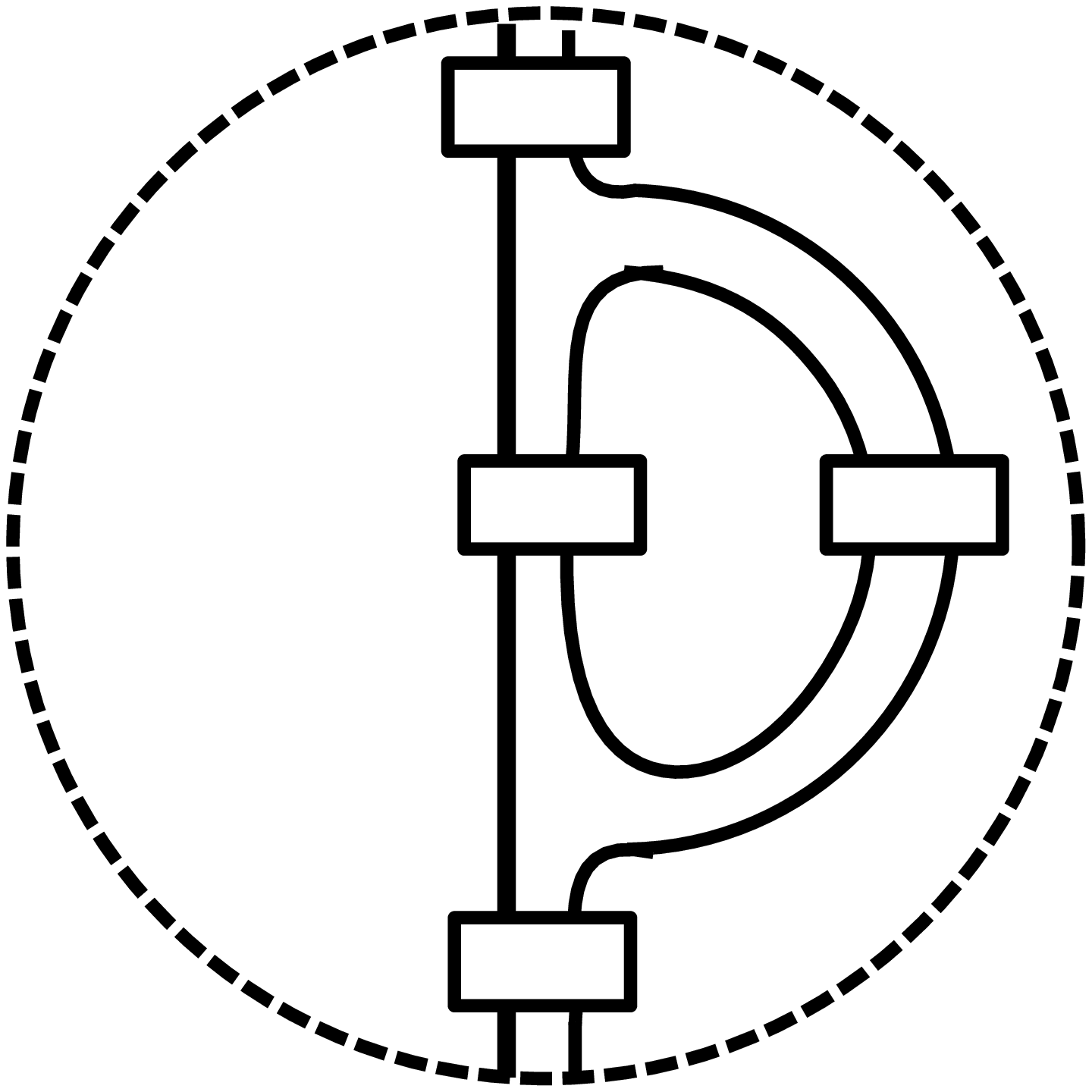}{45}{-15}{0}{0}.
\end{eqnarray}
The first line defines the area operator supported on  a framed
loop with vanishing self linking, represented here as an edge labeled by
2.  The second line expresses the first in the explicit
form of the vertex given in Fig.
(\ref{trivalent}b).  We may then use the identity of Eq. (44) to continue
\begin{equation}
8 \left(\hat{A}^q \right)^2 \ket{\Gamma^q \, n} = n^2
\pic{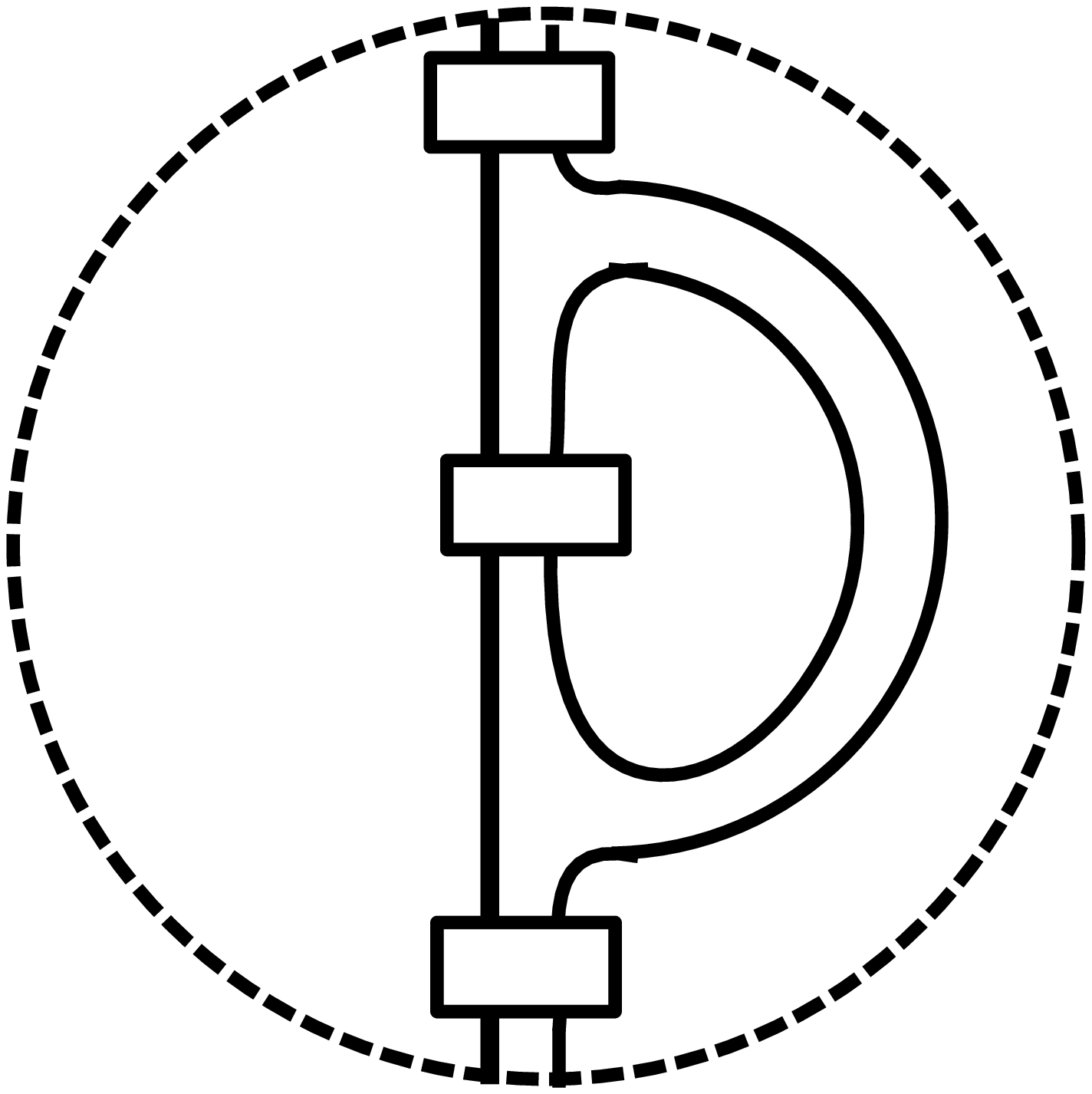}{45}{-15}{0}{0} -
{1 \over d} \pic{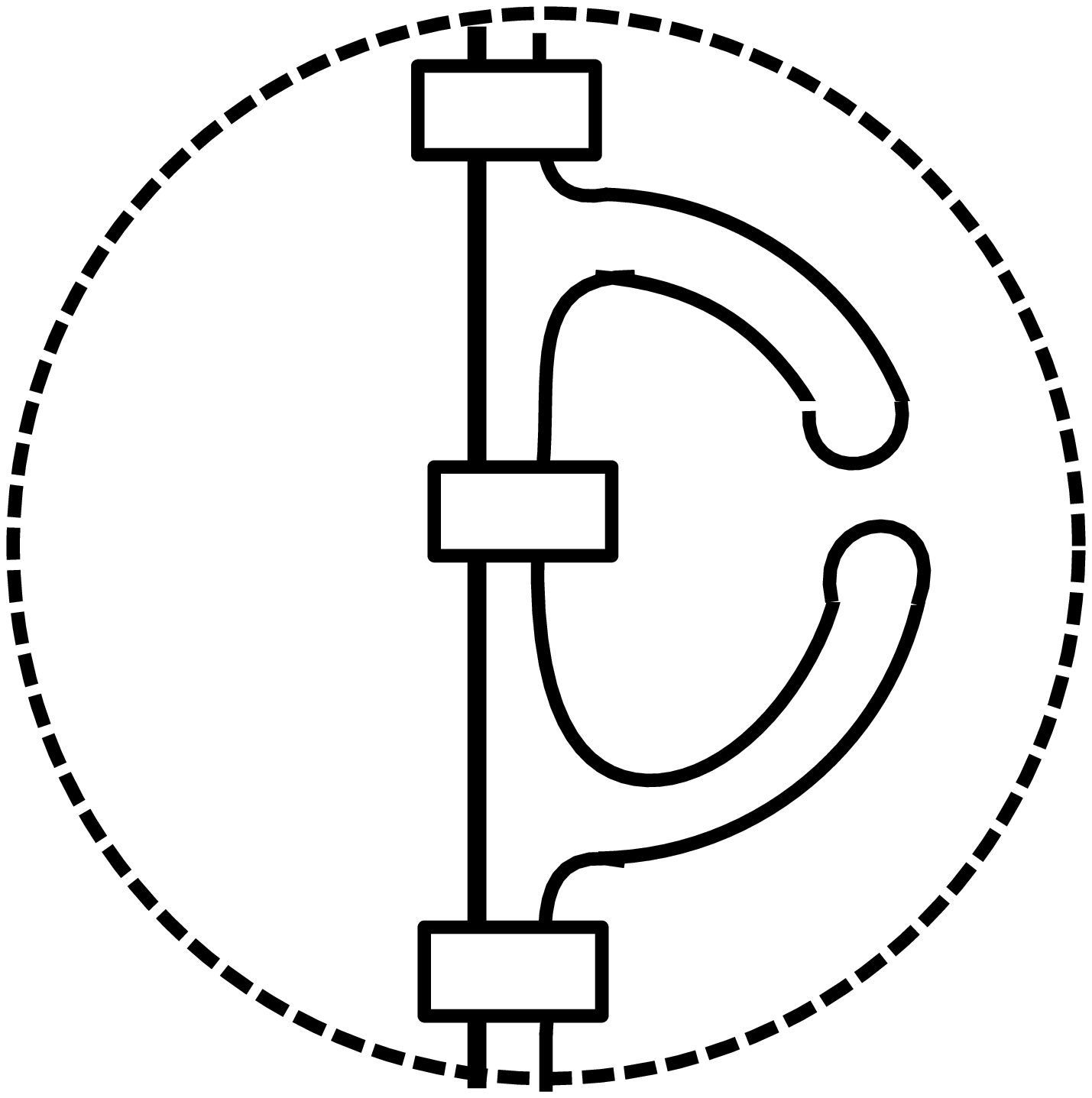}{45}{-15}{0}{0}
\end{equation}
which, using identity (\ref{closedloops})
reduces to, with constants restored,
\f
\hat{A}^q \ket{\Gamma^q \, n}  = l_{P}^2 { n \over 2} \sqrt{ {1 \over 2}
\left| {  [n+1] \over [n] } - {1 \over [2]} \right| } \ket{\Gamma^q \, n} .
\ff
One can easily verify that in the limit $A \rightarrow -1$
the usual eigenvalues proportional to $\sqrt{j(j+1)}$ (where $j=n/2$)
are recovered.
We may note that this result is not equal to the square root of the
$q$-deformed Casimir operator $[j][j+1]$.

\section{Eigenstates of the $\hat{T}_q[\alpha ]$}

q-spin nets have an interesting property for $q$ at a root
of unity, which is that there are only a finite number of representations
possible on each edge of a graph.  We can exploit this fact
to arrive at finite expressions for eigenstates of the
$\hat{T}_q[\alpha ]$ operators. This may allow us to define an
inverse transform that will enable us to define a notion that
corresponds to the conjugacy classes of connections in the
$q$-deformed case.

We consider an eigenstate of $\hat{T}_q[\alpha ]$ associated with
a simple, un-twisted unknot  $\tilde{\alpha}^f$.  This will be of the form
\f
\bra{\alpha} = \sum_{i=1}^k c_i \bra{\alpha , i}
\ff
where $\bra{\alpha,  i }$ is the spin network state associated to the
$i$ representation traced on the framed loop $\tilde{\alpha}^f$.
We want to find
the coefficients $c_i $ such that
\f
\bra{\alpha} \hat{T}_q[\alpha ] = \lambda \bra{\alpha}  .
\ff
We may use the identity shown in Eq. (\ref{edgeaddition}) to find,
\f
\bra{\alpha,  i } \hat{T}_q[\alpha ] = \bra{\alpha,  i+1} +
\bra{\alpha ,  i-1 }
\ff
where we use implicitly also that $\bra{\alpha , r-1 } =0$.
That is, the solution for the eigenstate uses crucially the
fact that the representations of $SU(2)_q$ extend only from
spin $0$ to spin $r-1$.
It is easy to extract the relations,
\f
c_{r-2} = \lambda c_{r-1} \ \ , \ \ \
\lambda c_{i}= c_{i+1}  +  c_{i-1} , \ \ \ 1 < i < r-1
\ff
\f
c_2 = \lambda c_1
\ff
These may be solved in all cases to find a polynomial in $\lambda$.
For a given $k$, and thus $r$, there are a finite number of eigenvalues, which
are given by solutions to
\f
\lambda W^{r-2} (\lambda ) = 1
\ff
where $W^i (\lambda )$ are defined by
\f
W^i (\lambda ) = \lambda - {1 \over W^{i-1} (\lambda )}
\ff
and
\f
W^1 (\lambda ) = \lambda.
\ff

\section{Discussion}

We close with a few comments on directions which
may be explored.  First, the framed multiloop
product $\cup$ may be extended to the general case including
arbitrary intersections.  This is done in \cite{BMS}.
The combinatorial argument for the
uniqueness and independence of the spin network basis that
we sketched above should be completed.  We expect that
this involves only a careful iteration of cases, as it is a simple
extension of known results about the Kauffman bracket.  This
would also be interesting as it would provide an alternative
proof of the independence
of the spin network basis even in the classical ($q=1$) case, which
would not rely on the connection representation.

There remains the question of whether
the formalism we have defined here represents a departure from the
notion that the state space of quantum gravity should be defined
in terms of measures.  It is interesting to conjecture that there may
be a completion of this space built on framed rather than ordinary loops.
Alternatively, there may exist a framework in
which an extended holonomies seen a maps from a framed loop
group (in the sense of Gambini and
collaborators\cite{gambini-looprep} to
$SU(2)_q$ or a related structure.  One way to such a construction
could be through an inverse
transform\cite{thomas-inverse} constructed using
the results of Section 6.

We may also remark that the structures we have described here
must bear some relationship to the general notion of extended
loops developed in \cite{extendedloops}.  The construction
of extended loops
was motivated by the similar considerations which gave rise to  framed
loops. They have the advantage that they are defined in terms of
functionals of connections and do allow a definition of integrals
related to Eq. (1).  However, examples are known in which framed loops
are not gauge invariant\cite{troy}, and thus are not well
defined on functionals
on ${\cal A}/{\cal G}$.  Although the question has not, to our
knowledge, been settled, it  is possible that there is a restricted
class of extended loops that are gauge invariant.  Such a notion
of restricted extended loops may be
related to the notion of framed loops that we have used here.

At the present it is not clear if such links between
the $q$-deformed loop algebra and the connection representation
will emerge. However, even if
the results of these investigations were negative, it would not mean
that the $q$-deformed loop representation is not useful for
quantum gravity.  Instead, it may be that
an aspect of the quantum world
expressible in terms of non-local observables based on framed loops is not
captured in terms of the classical description based on connections.
As the quantum world is
prior to its classical approximation, this
may reflect only the necessity of leaving behind the fiction of
deriving a quantum theory from its classical limit.

Finally, we may note that the $q$-deformed loop
representation may have
practical value in calculations in quantum gravity.  As will
be described elsewhere, the most efficient procedure for
computing with spin network states, which employs the
recoupling theory extends to the $q$-deformed
case\cite{inpreparation, KL}.  The main difference is that
because of the
restriction to $j \leq r-1$, one cannot concentrate more than
a fixed amount of
area on the edge of one graph, or too much volume on a vertex
 of a graph.  This means that for fixed $k$ the infinite
volume limit must be a limit in which graphs become larger and
more complex.  This may mean that both perturbative and
path integral calculations at finite $k$ may be better behaved
with respect to possible infrared divergences than the classical
$q=1$ case.   Even if the limit of large $k$, and hence
small cosmological constant is to taken in the end, the $q$-deformation
may then serve as a natural, diffeomorphism invariant infrared
regulator for non-perturbative quantum gravity.

\section*{ACKNOWLEDGEMENTS}

We would especially like to
thank Roumen Borissov and Carlo Rovelli for constant
encouragement and advice as well as for comments on earlier drafts of this
paper.  We are also
indebted to A. Ashtekar, J. Baez, L. Crane, L. Kauffman,
and J. Pullin for encouragement and critical comments.  One of us (LS)
would also like to thank the Institute for Advanced Study in Princeton and
SISSA in Trieste for hospitality while this work was carried out. This work
was supported, in part, by NSF grant number PHY-93-96246 to the Pennsylvania
State University.

\end{document}